\documentclass[aps,pra,10pt,twocolumn,a4paper,nofootinbib,preprintnumbers]{revtex4-1} 

\usepackage{Paper_style}

        \newcommand{\methods}{Methods}
        \newcommand{\supp}{Supplementary Materials}

                \newcommand{\AtoBLength}{253.9} %
                \newcommand{\AtoBLengthApprox}{254} %
                \newcommand{\AtoBLossLow}{56.0} 
                \newcommand{\AtoBLossLowPlusTwentydB}{76.0} 
                \newcommand{\AtoBAttCoeff}{0.22} 
            
                \newcommand{\AtoCLength}{156.7} 
                \newcommand{\AtoCLossHigh}{33.5} 
                \newcommand{\AtoCLossLow}{32.4} 
                
                \newcommand{\BtoCLength}{97.2} 
                \newcommand{\BtoCLossHigh}{22.2} 
                \newcommand{\BtoCLossLow}{21.7} 

            \newcommand{\PhaseDriftInLab}{9197} 
            \newcommand{\PhaseDriftInField}{8256} 

            \newcommand{\PhaseDriftCoarseStab}{2.2} 
            \newcommand{\PhaseDriftFineStab}{0.47} 

            \newcommand{\PhaseStabUncertainty}{0.1} 

        \newcommand{\LRefWavelength}{1551.72} 
        \newcommand{\LQuantWavelength}{1550.12} 

        \newcommand{\FreeDriftLQInterfVis}{99.09} 
        \newcommand{\PartStabilisedLQInterfVis}{97.15} 

        \newcommand{\DeadTimeWindow}{5} 
        \newcommand{\ExtinctionRatioSandN}{35} 
        \newcommand{\PulsesDuration}{120} 
        \newcommand{\PatternRepRate}{1} 
        \newcommand{\EncodedPatternLength}{418800} 

        \newcommand{\SysRefClock}{125} 

        \newcommand{\ProtExecTime}{7.5} 
            \newcommand{\ExpSKRBitPerPulse}{2.20E-7} 
            \newcommand{\ExpSKRBitPerSecond}{110.1} 
            \newcommand{\ExpSKRBitPerSecondAbstract}{110} 
            
            \newcommand{\AvgQberXBasisU}{4.61} 
        
            \newcommand{\AvgQberXBasisV}{5.84} 

\newacronym{aopp}{AOPP}{Actively Odd-Parity Pairing}
\newacronym{apd}{APD}{Avalanche Photo Diode}
\newacronym{awg}{AWG}{Arbitrary Waveform Generator}
\newacronym{bs}{BS}{Beam Splitter}
\newacronym{cir}{CIR}{Circulator}
\newacronym{cw}{CW}{Continuous Waveform}
\newacronym{cwdm}{CWDM}{Coarse Wavelength Division Multiplexing}
\newacronym{dc}{DC}{Direct Current}
\newacronym{dlcz}{DLCZ}{Duan-Lukin-Cirac-Zoller}
\newacronym{dwdm}{DWDM}{Dense Wavelength Division Multiplexing}
\newacronym{edfa}{EDFA}{Erbium Doped Fibre Amplifier}
\newacronym{epc}{EPC}{Electronic Polarisation Controller}
\newacronym{fpga}{FPGA}{Field-Programmable Gate Array}
\newacronym{im}{IM}{Intensity Modulator}
\newacronym{ld}{LD}{Laser Diode}
\newacronym{oil}{OIL}{Optical Injection Locking}
\newacronym{otdr}{OTDR}{Optical Time-Domain Reflectometry}
\newacronym{pbs}{PBS}{Polarisation Beam Splitter}
\newacronym{pm}{PM}{Phase Modulator}
\newacronym{pmp}{PMP}{Post-Measurement Pairing}
\newacronym{pmpqkd}{PMP-QKD}{Post-Measurement Pairing QKD}
\newacronym{qber}{QBER}{Quantum Bit Error Rate}
\newacronym{qc}{QC}{Quantum Communications}
\newacronym{qkd}{QKD}{Quantum Key Distribution}
\newacronym{rf}{RF}{Radio Frequency}
\newacronym{snspd}{SNSPD}{Superconductive Nanowire Single Photon Detectors}
\newacronym{skc0}{SKC\textsubscript{0}}{Repeaterless Secret Key Capacity bound}
\newacronym{skc1}{SKC\textsubscript{1}}{Single Repeater Secret Key Capacity bound}
\newacronym{skr}{SKR}{Secret Key Rate}
\newacronym{smf}{SMF}{Single Mode Fibre}
\newacronym{tf}{TF}{Twin Field}
\newacronym{tfqkd}{TF-QKD}{Twin Field QKD}
\newacronym{voa}{VOA}{Variable Optical Attenuator}
\newacronym{wcp}{WCP}{Weak Coherent Pulse}

\begin{document}

\graphicspath{ {./images/} }


\title{Coherent Quantum Communications Across National Scale Telecommunication Infrastructure}

\author{
Mirko Pittaluga$^{1 \ast}$,
Yuen San Lo$^{1}$,
Adam Brzosko$^{1}$,
Robert I. Woodward$^{1}$,
Matthew S. Winnel$^{1}$,
Thomas Roger$^{1}$,
James F. Dynes$^{1}$,
Kim A. Owen$^{1}$,
Sergio Ju\'arez$^{1}$,
Piotr Rydlichowski$^{2}$,
Domenico Vicinanza$^{3,4}$,
Guy Roberts$^{3}$
\& 
Andrew J. Shields$^{1}$
}

\affiliation{\small
$^{1}$Toshiba Europe Limited, 208 Cambridge Science Park, Cambridge CB4 0GZ, UK\\
$^{2}$Poznan Supercomputing and Networking Center, Jana Paw\l{}a II 10, 61-139 Poznan, PL\\
$^{3}$G\'EANT Vereniging, Hoekenrode 3, 1102BR Amsterdam, NL\\
$^{4}$School of Computing and Information Science, Anglia Ruskin University, Cambridge, UK\\
$^{\ast}$mirko.pittaluga@toshiba.eu\\
}

\begin{abstract}
    Quantum communications harness quantum phenomena like superposition and entanglement to enhance information transfer between remote nodes.
    Coherent quantum communications, essential for phase-based quantum internet architecture, require optical coherence among nodes and typically involve single-photon interference.  
    Challenges like preserving optical coherence and integrating advanced single-photon detectors have impeded their deployment in existing telecommunication networks.
    This study introduces innovative approaches to the architecture and techniques supporting coherent quantum communications, marking their first successful integration within a commercial telecom infrastructure between Frankfurt and Kehl, Germany.
    Employing the Twin Field Quantum Key Distribution protocol, we achieved encryption key distribution at \qty{\ExpSKRBitPerSecondAbstract}{\bit\per\second} over \qty{\AtoBLengthApprox}{\kilo\metre}.
    This system features measurement-device-independent properties and non-cryogenically cooled detectors, and represents the first effective quantum repeater implementation on telecom infrastructure, the longest practical quantum key distribution deployment to date, and validates the feasibility of a phase-based quantum internet architecture.
\end{abstract}

\maketitle

\section*{Introduction}

\gls*{qc} exploit the properties of quantum mechanical states to enable applications that are provably impossible using classical\footnote{
    In this context, the term \textit{classical} refers to all communication and computing processes that can be implemented without relying on quantum mechanical phenomena.
    } 
means alone.
These applications include information-theoretic secure communication~\cite{BB84, Ekert.1991, Xu.2020b, Pirandola.2020}, exponential speed-up in communication~\cite{Buhrman.2001, Buhrman.2010, Arrazola.2014}, quantum sensor networks~\cite{Gottesman.2012, Komar.2014, Degen.2017, Liu.2021a, Malia.2022}, networked quantum computing~\cite{Kimble.2008, Tani.2005, Denchev.2008, Barz.2012, BenOr.05222005}, and more~\cite{Pfaff.2014, Northup.2014, Hensen.2015, Roberts.2017c, HerreroCollantes.2017}.
\acrfull*{qkd} is the most developed of these, providing information-theoretic secure key distribution over untrusted public channels without algorithmic assumptions and regardless of the resources and capabilities of potential attackers.
Due to their long-distance transmission capability, photons are the preferred information carriers for \gls*{qc} and are typically encoded as qubits, the elemental unit of quantum information.
When \gls*{qc} connect multiple quantum-capable nodes, a quantum network is formed, or, on a larger scale, a ``quantum internet''. 

Coherent \gls*{qc}, a subset of \gls*{qc}, requires maintaining optical coherence between nodes.
Optical coherence is a powerful resource, supporting various \gls*{qc} applications~\cite{Gottesman.2012, Arrazola.2014}.
Notably, it underpins the \gls*{dlcz} teleportation mechanism~\cite{Duan.2001}, which is one of the most practical implementations of a fundamental primitive for the quantum internet~\cite{Bennett.1993}.
This mechanism has enabled the first 3-node quantum networks~\cite{Pompili.2021, Liu.01092023} and quantum teleportation across non-adjacent nodes of a network~\cite{Hermans.2022}.

When used in \gls*{qkd}, optical coherence facilitates protocols like \gls*{tfqkd}~\cite{Lucamarini.2018, Wang.2018, Curty.2019} and \gls*{pmpqkd}~\cite{Zeng.2022, Xie.2022, Avesani.2023}, which are the most practical ways to overcome the \gls*{skc0}\cite{Pirandola.2017}, the fundamental capacity of a quantum channel not endowed by a quantum repeater~\cite{Pirandola.2020}.
They also double the maximum communication distance compared to standard point-to-point \gls*{qkd} protocols, expanding the reach of fibre-based \gls*{qkd} to national and international distances.
Several \gls*{tfqkd} lab demonstrations~\cite{Minder.2019} have achieved distances exceeding hundreds of kilometers~\cite{Pittaluga.2021, Liu.2021, Chen.2021, Chen.2022, Wang.2022, Clivati.2022}, with one reaching up to \qty{1000}{\kilo\metre}~\cite{Liu.2023}.

\gls*{qc} depends on appropriate channels and nodes to transmit and process quantum information.
The widespread adoption of the quantum internet is contingent on aligning \gls*{qc} implementation requirements and practices with the capabilities of existing telecommunication infrastructures. 
Presently, integration has only been achieved for simpler, point-to-point, non-coherent \gls*{qkd} systems.
Coherence-based \gls*{qc} protocols have been primarily tested in laboratories with specialised equipment like ultra-stable optical cavities and cryogenic photon detectors, which are unsuitable for typical telecommunication environments. 
Given the benefits of using optical coherence for \gls*{qc}, developing methods to integrate it with the existing telecommunication framework is crucial for the field advancement and large-scale adoption.

This study introduces the first integration of coherent \gls*{qc} into commercial telecommunication infrastructure, linking colocation data centres in Germany via optical fibres with high losses and asymmetric link lengths.
The system implemented the \gls*{tfqkd} protocol and incorporated practical technologies that facilitate seamless operation.

Implementing coherent, or phase-based, \gls*{qc} poses several challenges such as establishing a common phase reference frame among distant encoding users and mitigating the phase noise stemming from lasers and transmission channels.
To address these challenges, we developed a practical architecture for optical frequency dissemination and orchestration among network nodes and active off-band phase stabilisation using non-cryogenically cooled detectors. 

In our system, the central node distributes two optical frequency references to the transmitting nodes through a service fibre, allowing them to lock their lasers to a common frequency reference and achieve mutual phase-locking. 
This eliminates laser phase noise more practically and cost-effectively than using ultra-stable lasers and external cavities~\cite{Chen.2021, Clivati.2022, Zhou.2023, Liu.2023}.
To mitigate phase noise from the fibre, we use an off-band phase stabilisation feedback system~\cite{Pittaluga.2021} based on single-photon interference outcomes monitored by \glspl*{apd}.
This combination of technologies is crucial for our experiment's success.

While \glspl*{apd} offer practical single-photon detection capabilities, they possess less ideal properties compared to \gls*{snspd}, the detection technology used to support coherent \gls*{qc} so far~\cite{Liu.2021, Chen.2021, Wang.2022, Liu.2023, Pompili.2021, Hermans.2022, Liu.01092023}.
As summarised in \cref{tab:APDs_vs_SNSPDs}, \glspl*{apd} exhibit higher dark counts, lower detection efficiency, and susceptibility to afterpulse effects.
However, \glspl*{apd} are one to two orders of magnitudes more economical than \glspl*{snspd}, are more practical, and can operate at temperatures compatible with telecommunication infrastructure.

\begin{table}[htbp]
    \centering
        \begin{tabular}{lcc}
            \hline
                                            & \textbf{APDs}                                             & \textbf{SNSPDs}                                           \\
            \hline
            Detection efficiency (\%)       & 10 - 20                                                   & 60 - 80                                                   \\
            Dark Counts                     & Hundreds Hz                                               & Few Hz                                                    \\
            Afterpulse  (\%)                & $\sim$3                                                   & 0                                                         \\
            Size (U)                        & $<$1                                                      & $>$10                                                       \\
            Cost (\$)                       & \num{1000}s                                               & \num{100000}s                                             \\
            Cooling system                  & Thermoelectric                                            & Cryogenic                                                 \\
                                            & \qtyrange[range-units = bracket]{+20}{-50}{\celsius}      & \qtyrange[range-units = bracket]{-270}{-272}{\celsius}    \\
            \hline
        \end{tabular}
    \caption{
    Comparison of typical properties of \acrshortpl*{apd} and \acrshortpl*{snspd}.
    The cost indicates the order of magnitude for an entire \acrshort*{apd} or \acrshort*{snspd} system.
    The system sizes are expressed in rack units (U), the typical unit of measure for telecom rack-mounted equipment.
    1 U corresponds to \qty{44.45}{\milli\metre} thickness.
    }\label{tab:APDs_vs_SNSPDs}
\end{table}

The success of this implementation hinges on combining the off-band stabilisation technique with \glspl*{apd}.
For long-distance coherent \gls*{qc}, distributing significantly brighter phase-reference signals than the protocol-encoded ones is crucial to establishing a common phase reference frame among users.
Previous methods, which used time-multiplexed phase-reference pulses at the same optical frequency as the encoded signals, are incompatible with \glspl*{apd} due to afterpulse effect~\cite{Hadfield.2009, Zhang.2015, Humer.2015}.
Using the same \glspl*{apd} to detect intense reference pulses and protocol-encoded signals introduces noise that obscures the encoded signal due to the afterpulse effect.
By contrast, our off-band stabilisation employs distinct optical frequencies for phase-reference and protocol-encoded signals, allowing separate detectors for each, eliminating afterpulse cross-contamination.

\section*{System Details and Methods}\label{sec:sys_details}
\begin{figure*}[ht!]
    \centering
    \includegraphics[width=\textwidth]{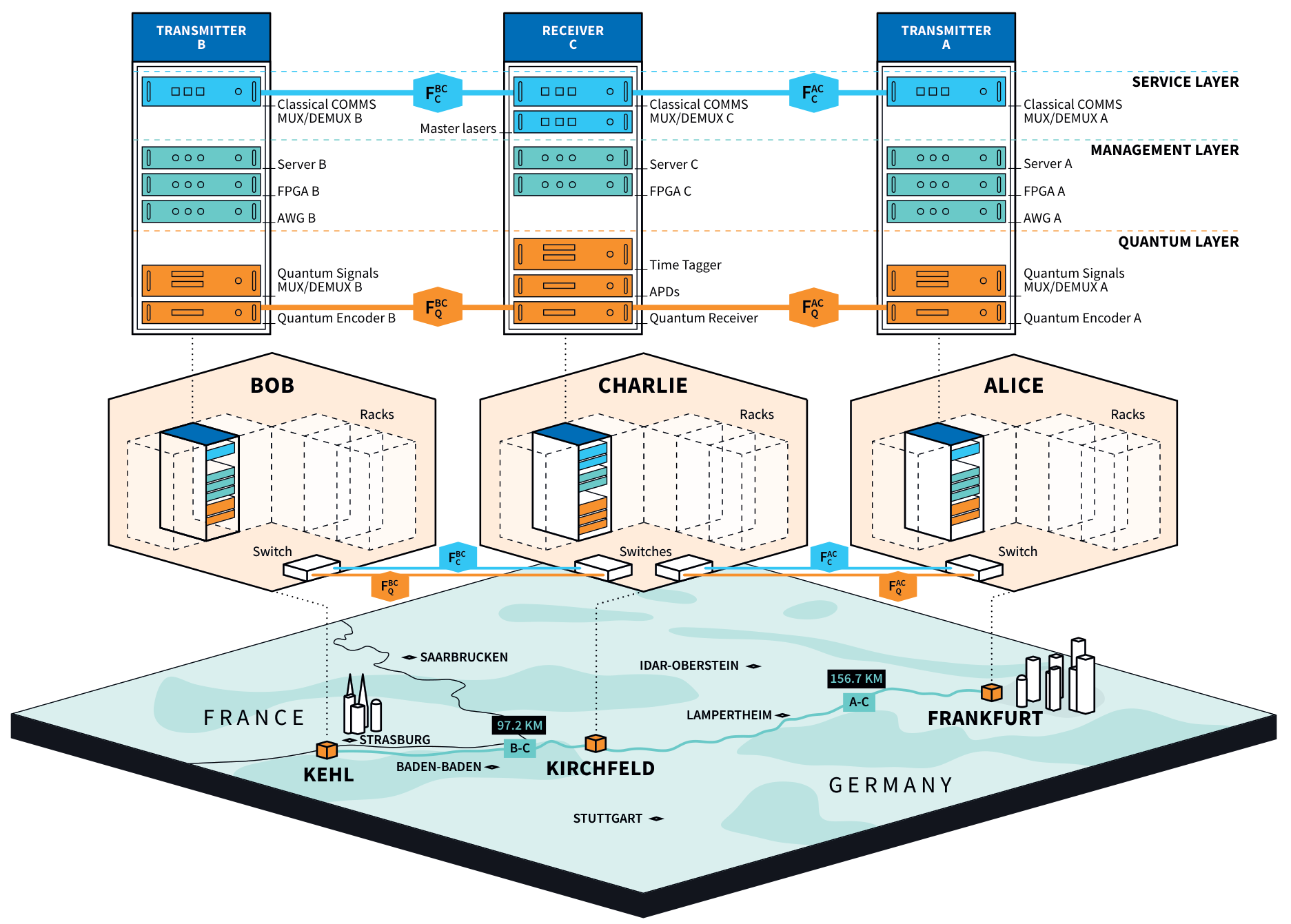}
    \caption{
    The \gls*{tfqkd} system consists of three nodes: two transmitter nodes, Alice and Bob, and one receiver node, Charlie (called A, B, and C respectively).
    Equipment for nodes A, B and C was installed in three colocation data centres located in Frankfurt, Kehl and Kirchfeld alongside other running telecommunication equipment.
    All equipment was mounted in 19-inch telecom racks, and the three nodes were interconnected pairwise through two fibre duplexes.
    The A-C fibre duplex spans \qty{\AtoCLength}{\kilo\metre}, with the two bundled fibres, $F^{AC}_Q$ and $F^{AC}_C$, characterised by losses of \qtylist{\AtoCLossLow;\AtoCLossHigh}{\decibel} respectively.
    The B-C duplex spans \qty{\BtoCLength}{\kilo\metre}, with $F^{BC}_Q$ and $F^{BC}_C$ characterised by losses of \qtylist{\BtoCLossLow;\BtoCLossHigh}{\decibel}, respectively.
    \textbf{Top:} Functional schematic of the subsystems installed in the three colocation centres, with equipment grouped into three layers spanning the three nodes based on their implemented functionality.
    }
    \label{fig:sys_installation}
\end{figure*}

The experiment was conducted in Germany, with access to the network infrastructure provided by G\'EANT.
Communication was established over a link spanning \qty{\AtoBLength}{\kilo\metre} between Frankfurt and Kehl with a \qty{\AtoBLossLow}{\decibel} loss and a relay in Kirchfeld, approximately three-fifths of the way. 
This setup forms a star-shaped quantum network with three nodes: two transmitters, Alice and Bob, at the network's edge, and a central relay receiver, Charlie.
Charlie is connected to each transmitter by a fibre duplex cable (characterisation is provided in \supp).
Equipment was housed within standard telecom racks in colocation data centres, functioning alongside existing telecommunications gear.
The network's geographical and functional layout, including the fibre links and system elements, is detailed in \cref{fig:sys_installation}.

The transmitters prepared optical \gls*{wcp} qubits for the quantum communication protocol, while the receiver node distributed common optical frequency references to ensure coherence and interfered the \glspl*{wcp} from transmitters.
Each node features a modular design with interconnected 19-inch rack boxes, enhancing compatibility with telecom racks and system flexibility.
The nodes' subsystems are organised into three functional layers: \textit{service}, \textit{management}, and \textit{quantum}, each spanning across the three nodes, as shown in the top diagram of \cref{fig:sys_installation}.

\begin{figure*}[ht!]
    \centering
    \includegraphics[width=\textwidth]{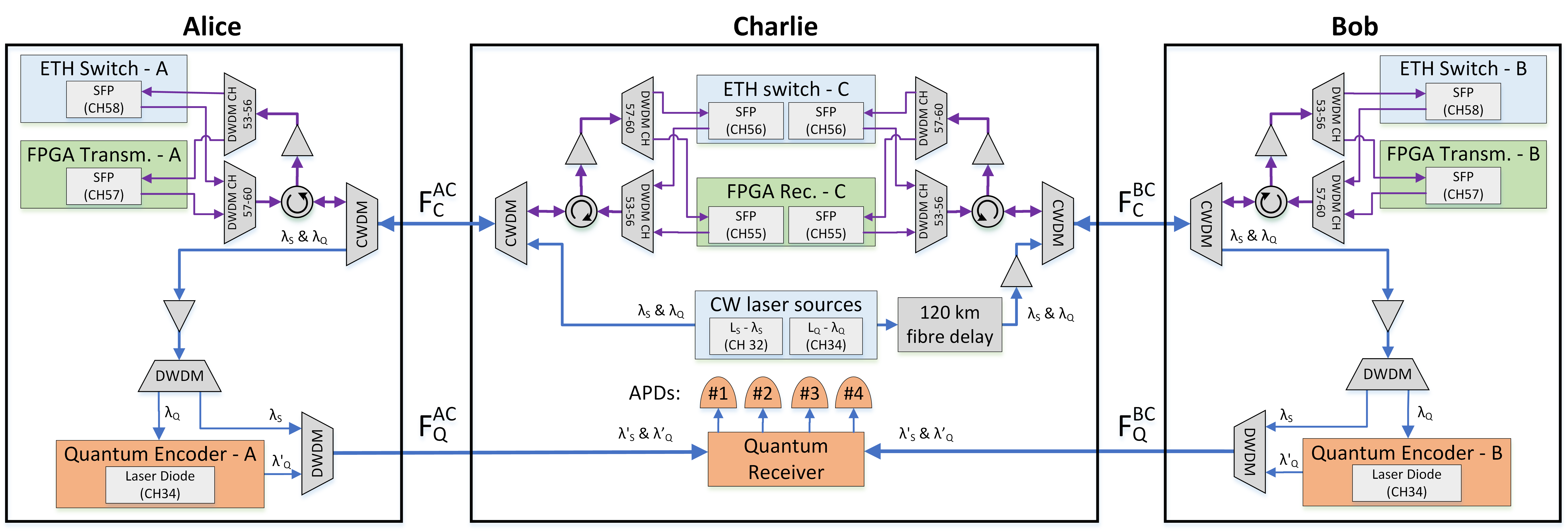}
    \caption{
    Routing of the optical signals in the quantum communication system.
    The service fibres $F^{AC}_C$ and $F^{BC}_C$ distribute the optical frequency references $\lambda_S$ and $\lambda_Q$ from the central to the transmitter nodes, while also supporting the bidirectional classical communication needed to establish a time-synchronised local network.
    $F^{AC}_Q$ and $F^{BC}_Q$ transmit the encoded $\lambda'_S$ and $\lambda'_Q$ signals from the transmitters to the central node, where they interfere.
    A comprehensive schematic of the system is reported in the \supp.
    Triangles: \acrshortpl*{edfa};
    Circles: circulators;
    Trapezoids: \acrshortpl*{cwdm} and \acrshortpl*{dwdm}.
    }
    \label{fig:sys_scheme}
\end{figure*}

The \textit{service layer} facilitates remote operation, intra-node communication, and distributes two optical frequency references ($\lambda_S$ and $\lambda_Q$) from Charlie, to the transmitters along with a clock for synchronisation of electronics.
It comprises, among other components, two master lasers ($L_S$ and $L_Q$), service fibres ($F_{C}^{AC}$ and $F_{C}^{BC}$), Ethernet switches equipped for optical networking, and a \gls*{dwdm}-based optical routing setup. 

The \textit{management layer} controls all system equipment.
It includes a server (Server\textsuperscript{A,B,C}) and a custom-programmed FPGA (FPGA\textsuperscript{A,B,C}) per node, and an \gls*{awg} for each transmitter.

A description of the components and functionalities implemented by the \textit{service} and \textit{management layers} is reported in the \methods, and a simplified representation of the routing of the optical signals in the system is shown in \cref{fig:sys_scheme}.
For a comprehensive schematic of the system refer to the \supp.

The \textit{quantum layer} executes the quantum communication protocol.
The specific protocol implemented in this experiment is the Sending-Not-Sending variant\cite{Wang.2018, Jiang.2020} of \gls*{tfqkd}\cite{Lucamarini.2018}.
This protocol enables continuous generation of a shared secret bit string between Alice and Bob, enabling quantum-safe communication by means of symmetric-key cryptography.
The transmitters encode information on the optical phase of \glspl*{wcp} which are sent to Charlie, where they interfere.
Charlie monitors the interference outcome with two single-photon detectors and announces the results (which detector clicked and when) on a public channel.
Charlie's measurements constitute a Bell state measurement of the joint quantum system composed of Alice's and Bob's \gls*{wcp} qubits.
By combining the publicly announced interference results with their private knowledge of the \glspl*{wcp} encoding, the transmitters initiate a sifting procedure (described in the \methods) that generates a shared secret bit string. 
The \gls*{tfqkd} protocol ensures measurement-device-independence for the detections made by Charlie~\cite{Lo.2012, Braunstein.2012}, and a favourable square-root scaling of the \gls*{skr} with channel loss~\cite{Lucamarini.2018}.
The transmitter nodes have two subsystems in the \textit{quantum layer}, the Quantum MUX/DEMUX Unit, and the Quantum Encoder.

The Quantum MUX/DEMUX Unit receives two optical frequency references distributed by Charlie ($\lambda_S$ and $\lambda_Q$) from the \textit{service layer}, separates them, and routes them appropriately for local processing and encoding before launching them into the quantum channel as $\lambda'_S$ and $\lambda'_Q$.

The Quantum Encoder employs \gls*{oil}~\cite{Ye.2003, Comandar.2016c, Minder.2019} to frequency-lock a locally controlled \gls*{ld} to the frequency reference $\lambda_Q$ distributed by Charlie.
The \gls*{ld}'s output is modulated by three \glspl*{im} and two \glspl*{pm} arranged in series, which encode \qty{\PulsesDuration}{\pico\second} long optical pulses at \qty{1}{\giga\hertz} with four possible optical intensities (required for the decoy-state method \cite{Lo.2005, Wang.2005}) and arbitrary phase in the $\left[0,2\pi\right)$ interval.
The encoded \glspl*{wcp} (also referred to as $\lambda'_Q$) are then adjusted for photon flux and polarisation and passed back to the Quantum MUX/DEMUX unit for transmission to the receiver.
Only for Bob, the $\lambda'_Q$ passes through a fibre stretcher before entering the Quantum MUX/DEMUX Unit.
The fibre stretcher serves as an actuator for a fine phase stabilisation mechanism that will be described below. 
Detailed operations of the Quantum MUX/DEMUX and Encoder units are outlined in the \supp.

Multiplexing the manipulated $\lambda'_S$ and $\lambda'_Q$ into the quantum channels is crucial for the off-band phase stabilisation technique~\cite{Pittaluga.2021}, which ensures complete phase stability of an optical channel in two steps: coarse stabilisation via a phase modulator at the receiver affecting both $\lambda'_S$ and $\lambda'_Q$, and fine stabilisation using a fibre stretcher in Bob, specifically targeting $\lambda'_Q$.
Together, these stabilisations maintain coherence between optical signals prepared at distant nodes.
This process is detailed further in the \methods.

The $\lambda'_S$ and $\lambda'_Q$ signals from Alice and Bob reach the receiver unit in Charlie via the quantum channels $F_{Q}^{AC}$ and $F_{Q}^{BC}$.
At the receiver, these signals are processed by Charlie's \textit{quantum layer} consisting of a quantum receiver, four \glspl*{apd}, and a time tagger. 
Initially, the polarisation of all received signals is aligned to Charlie's preferred axes.
Bob's signals undergo coarse phase stabilisation via a LiNbO\textsubscript{3} phase modulator, and then $\lambda'^{A}$ and $\lambda'^{B}$ signals interfere on a 50/50 \gls*{bs}. 
The interference outcomes, separated by two \glspl*{dwdm}, are monitored by three \qty{-30}{\celsius}-cooled \qty{1}{\giga\hertz} self-differenced \glspl*{apd}~\cite{Yuan.2008} (\cref{fig:sys_scheme}).
APD\textsubscript{1} and APD\textsubscript{2} monitor the outputs of the $\lambda'_Q$ signals interference, and their clicks are collected by a time tagger which records the time-of-arrival of each photon detection, required for the protocol execution.
APD\textsubscript{3} monitors the $\lambda'_S$ signal output, with clicks recorded by this detector utilised to aid the coarse feedback of the off-band phase stabilisation.

The protocol execution relies on several automatic stabilisation mechanisms within the \textit{quantum layer}.
These mechanisms optimise \gls*{oil} in the Quantum Encoders, ensure appropriate single photon fluxes at the transmitters, compensate for polarisation fluctuations introduced by the quantum channels, correct optical phase noise from the channels, and align the timing of encoded signals from Alice and Bob with Charlie’s detection gates.
Additional details on these mechanisms are available in the \supp.
Notably, the quantum communication system is designed for remote control and autonomous operation in physically restricted locations to comply with policies of modern deployed telecom infrastructure.

\section*{Results}

\begin{figure}[t!]
    \centering
    \includegraphics[width=1\columnwidth]{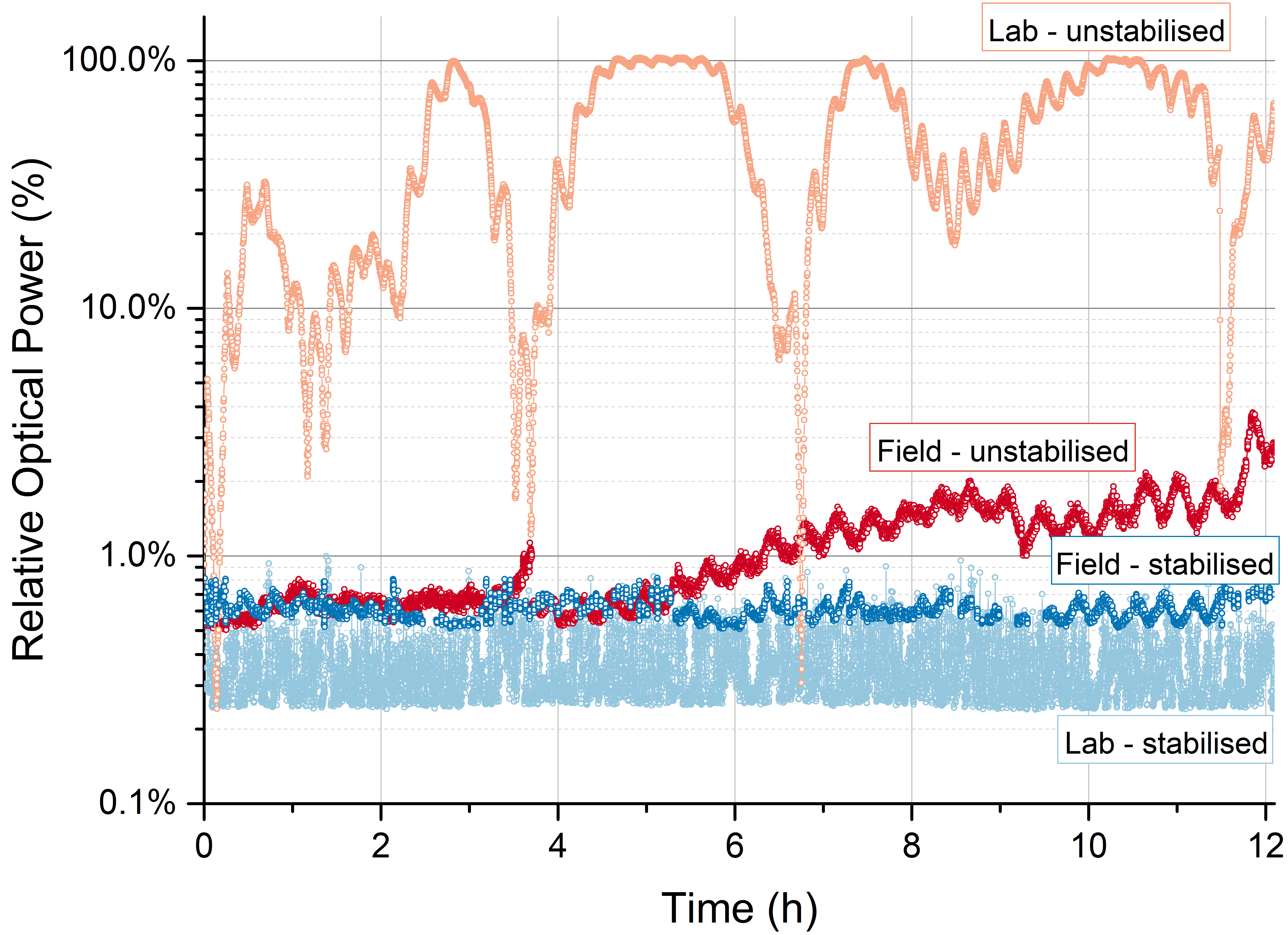}
    \caption{
    Comparison of optical intensities (relative to their maximum values) associated with polarisation rotations introduced by the channels. 
    The graph illustrates the orthogonal components of $\lambda'^{A}_{S}$ recorded in the lab or the field by APD\textsubscript{4}.
    These signals travel over a \qty{313.4}{\kilo\metre} long channel, equivalent to twice the A-C distance due to the looping-back of the support signal by the transmitters.
    The effect of polarisation rotation drift measured in the lab (pink) significantly exceeds that measured in the field (red).
    Spontaneous polarisation rotations introduced by the installed fibres resulted in an intensity drift of the orthogonal polarisation component from its minimum value to only 4\% of its maximum value over \qty{12}{\hour}.
    In the lab testbed, the same signal fluctuated four times between its minimum and maximum values over the same period.
    Upon activation of the polarisation stabilisation mechanism, it effectively maintains the intensity of the orthogonal polarisation below 0.8\% of its maximum value both in the lab (cyan) and in the field (blue).
    }
    \label{fig:pol_stability}
\end{figure}

Before the field experiment, we conducted laboratory tests using a testbed that mimicked the actual field conditions with \gls*{smf} spools and fixed optical attenuators to emulate the distance and channel losses of the installed fibres.
These tests were instrumental for the system optimisation and provided data acquired in a controlled environment for comparison with field results.

After deploying the system in Germany, we monitored the polarisation drift of the installed optical fibres by assessing the optical intensity of the orthogonal polarisation component recorded by APD\textsubscript{4}.
Field data showed higher polarisation stability compared to the lab environment, with minimal intensity drift of the orthogonal polarisation component over \qty{12}{\hour}, attributed to the natural temperature stability of underground fibres.
In contrast, lab-installed fibres, despite being in a temperature-controlled environment, exhibited more fluctuation due to less consistent temperature conditions. 
This is observed for all four signals entering the quantum receiver and is illustrated in \cref{fig:pol_stability} for $\lambda'^{A}_{S}$. 
Our tests also validate our polarisation stabilisation mechanism, which consistently maintains a \qty{21}{\decibel} intensity contrast between the preferred and its orthogonal polarisation axes.

In our coherent quantum communication system, maintaining coherence among quantum states encoded by different users is crucial for supporting phase-based protocols and minimising errors.
Sustained high-visibility optical interference between two optical signals is a reliable metric for high-quality coherence distribution.
In our case, the most relevant interference is between $\lambda'_Q$ signals from Alice and Bob, as detected by Charlie.

To ensure system coherence, we first evaluated the intrinsic coherence of our laser sources, $L_Q$ and $L_S$ (\cref{fig:sys_scheme}).
Due to significant link length discrepancies between Charlie to Alice and Bob, we introduced a \qty{120}{km} long delay fibre in Charlie's $\lambda'_S$ and  $\lambda_Q$ signals path to Bob, effectively reducing the optical path mismatch to about \qty{1}{km} --- well within the coherence length of $L_Q$ and $L_S$, which have a linewidth of a few hundred \unit{\hertz}.
More details on this aspect are reported in the \methods.

Additionally, we addressed signal degradation caused by the passage through optical components, amplification noise from \glspl*{edfa}, suboptimal \gls*{oil}, and cross-talk from classical signals in the \textit{service layer}.
These issues were mitigated by continuous \gls*{oil} optimisation at the transmitters, careful \gls*{edfa} selection, and narrow spectral filtering of $\lambda_S$ and $\lambda_Q$ at the transmitters, ensuring maximal interference visibility at the receiver.
As a result, the interference visibility between $\lambda'^A_Q$ and $\lambda'^B_Q$ free-drifting signals is \qty{\FreeDriftLQInterfVis}{\percent}, indicating robust coherence maintenance of our complex setup.

Furthermore, when coarse stabilisation is active, interference visibility drops slightly due to noise from $\lambda'_S$ signals and phase noise from coarse feedback, measuring at \qty{\PartStabilisedLQInterfVis}{\percent} in the field.
This visibility supports effective coherent quantum communication with a low error rate.

In phase-based quantum communications, it is crucial to counterbalance the phase noise introduced by optical fibres for successful protocol execution.
The standard deviation of the phase drift, characterising this noise, is measured in the field as $\sigma_{UN}^{field}$=\qty{\PhaseDriftInField}{\radian\per\second} and $\sigma_{UN}^{lab}$=\qty{\PhaseDriftInLab}{\radian\per\second} in the lab.
These measurements indicate similar phase behaviours in both environments, with further details provided in the \supp.

\begin{figure}[ht!]
    \centering
    \includegraphics[width=1\columnwidth]{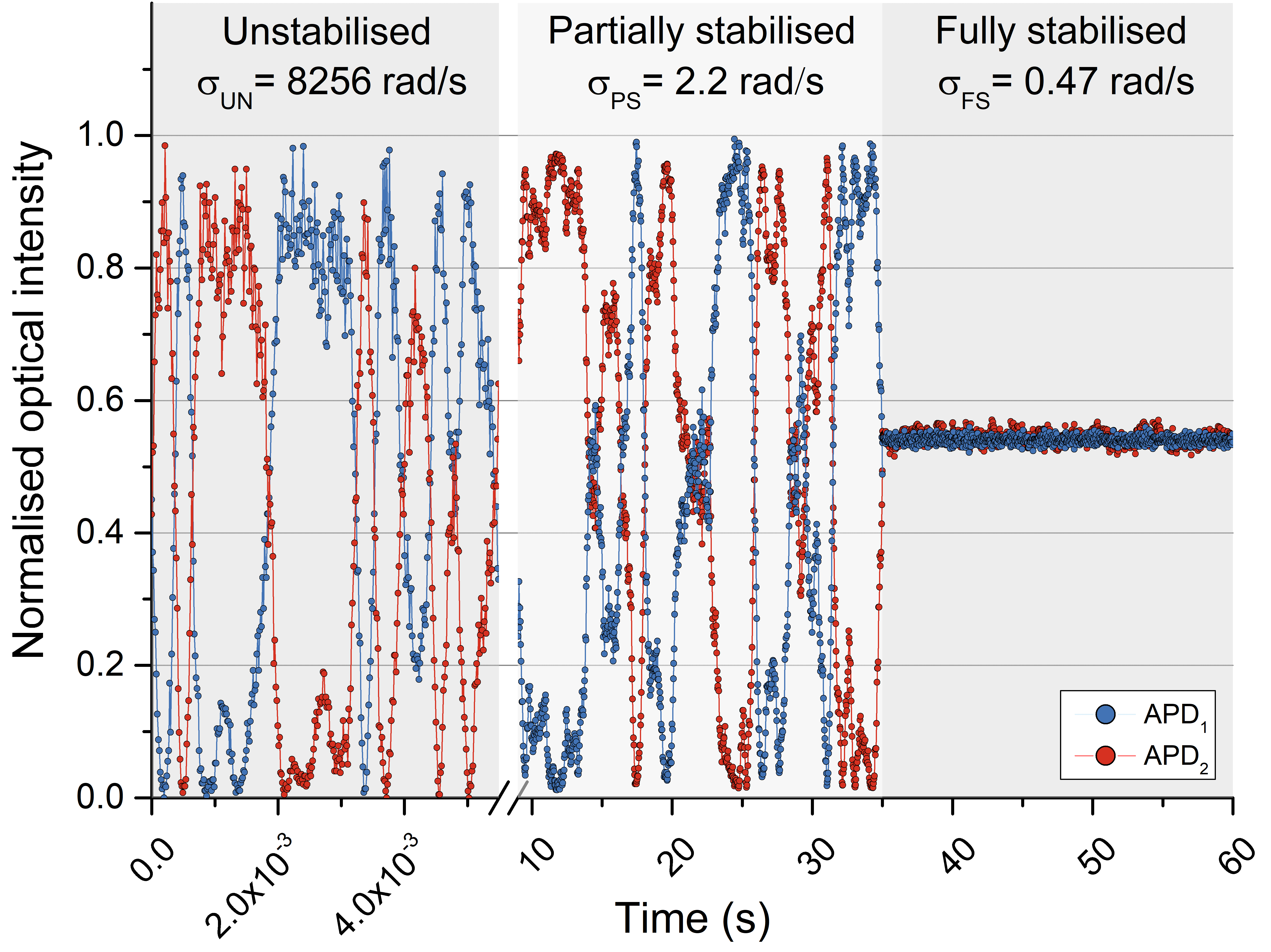}
    \caption{
    Optical interference between $\lambda'^{A}_{Q}$ and $\lambda'^{B}_{Q}$ measured by Charlie through APD\textsubscript{1} and APD\textsubscript{2} at three stages of the phase stabilisation.
    During the \emph{unstabilised} stage, the free-drifting interference between the signals is recorded.
    In the \emph{partially stabilised} stage, data is captured when only coarse phase stabilisation is activated.
    \emph{Full stabilisation} occurs when both coarse and fine phase stabilisations are activated.
    To accurately interpret the results, notice the interruption in the abscissa and the different time scales used.
    The increasing stability of the signal across the three stages can be characterised by the decrease in the average standard deviation of the phase drift, reported as the $\sigma$ values for each stage.
    }
    \label{fig:phase_stabilisation}
\end{figure}

The phase noise from the fibres is mitigated using an off-band phase stabilisation mechanism.
\Cref{fig:phase_stabilisation} shows how the interference between $\lambda'_{Q}$ signals evolves through different stabilisation stages.
Initially, with no stabilisation activated, significant intensity fluctuations occur.
Upon activating coarse phase stabilisation, the phase drift is significantly reduced to $\sigma_{PS}$=\PhaseDriftCoarseStab~\unit{\radian\per\second}.
Full stabilisation is achieved with fine stabilisation activated over $\lambda'_{Q}$, reducing the phase drift to $\sigma_{PS}$=\PhaseDriftFineStab~\unit{\radian\per\second} and locking the phase difference between the two arms to a target value with an uncertainty below \PhaseStabUncertainty~\unit{\radian}.

Minimising background noise detected by APD\textsubscript{1} and APD\textsubscript{2} is essential for successful protocol execution in our quantum communication system.
Key noise sources include \gls*{apd} dark counts, afterpulse effects, and suboptimal vacuum state encoding.
The dark count rate and afterpulse effect are an inherent property of the detectors that depend on their structure and operation conditions. 
Differently from the dark counts, the detrimental impact of the afterpulse on the key generation can be significantly mitigated by setting a long deadtime window to disregard subsequent counts following an initial successful detection \cite{Zhang.2010, Humer.2015}.
In this experiment, we set a deadtime window of \DeadTimeWindow~\unit{\micro\second}, eliminating the impact of afterpulseing on key generation.
To achieve a high \qty{\ExtinctionRatioSandN}{\decibel} intensity contrast between the brightest and dimmest encoded pulses, we employ three \glspl*{im} in series in the transmitters.

We used the developed system to implement the AOPP SNS TF-QKD protocol~\cite{Jiang.2020}, encoding repeated \qty{\PatternRepRate}{\giga\hertz} \glspl*{wcp} patterns.
Pulses were interleaved with a \qty{50}{\percent} duty cycle, alternating between protocol-encoded pulses for key generation and phase-unmodulated pulses for fine phase stabilisation.
Protocol-encoded pulses were prepared in two bases.
The $\mathbb{Z}$ (coding) basis includes phase-randomised pulses of two intensities, $s$ and $n$, representing the \textit{sending} and \textit{not-sending} encoding bits of the protocol.
The $\mathbb{X}$ (testing) basis includes phase-encoded pulses of three levels, $u$, $v$, and $w$, and is used to implement the decoy-state method~\cite{Lo.2005, Wang.2005}.
Pulse intensities were independently set for $s$, $u$, and $v$, while $n$ and $w$ were matched, yielding four intensity settings per transmitter.
Pulse intensities and distribution were optimised to maximise key generation in asymmetric link conditions, and to meet crucial security conditions for asymmetric protocol use~\cite{Hu.2019}.

The system achieved a \gls*{skr} of \qty{\ExpSKRBitPerPulse}{\bit\per\pulse}, equivalent to \qty{\ExpSKRBitPerSecond}{\bit\per\second} over \ProtExecTime~\unit{\hour} of uninterrupted operation. 
Over this period, the \gls*{qber} for the $\mathbb{X}$ basis averaged \AvgQberXBasisU~\% for $u$ pulses and \AvgQberXBasisV~\% for $v$ pulses. For a comprehensive patterns and results discussion description, refer to the \supp.

\begin{figure*}[ht!]
    \centering
    \includegraphics[width=0.95\textwidth]{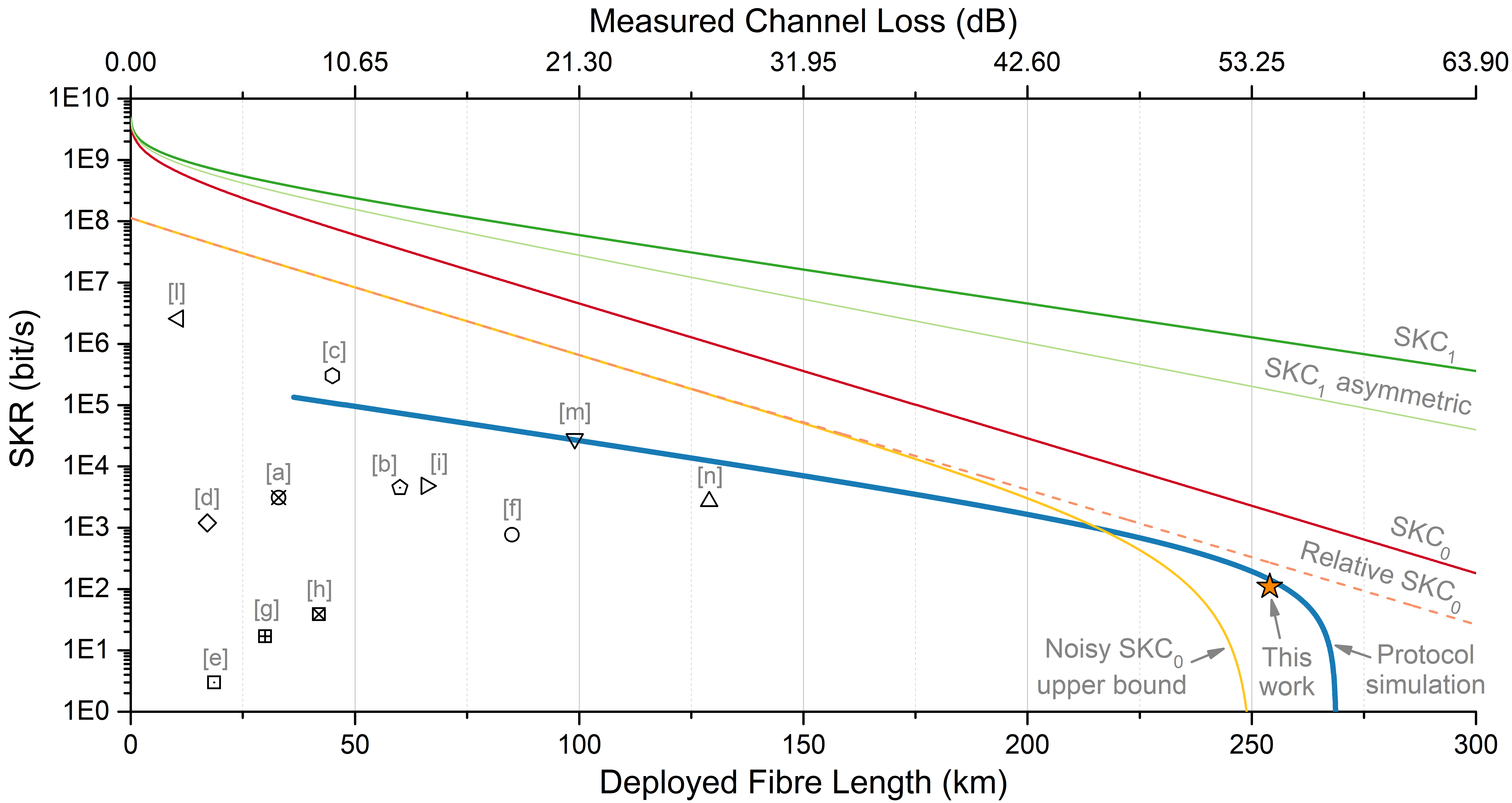}
    \caption{
    The experimentally obtained \gls*{skr} (star) and the simulation for our protocol (blue line) are plotted against the quantum channel length (bottom) and attenuation (top).
    The channel attenuation is calculated using the \qty{\AtoBAttCoeff}{\decibel \per \kilo \metre} attenuation coefficient measured for this experiment.
    Other lines represent theoretical capacity bounds of quantum communication schemes, with analytical expressions provided in the \methods.
    \acrshort*{skc1} is the bound for symmetric quantum communications aided by a single quantum repeater~\cite{Pirandola.2017}, while \acrshort*{skc1} asymmetric represents the same bound for a channel where the repeater is positioned three-fifths of the way between the transmitters, as in our case.
    \gls*{skc0} is the repeaterless secret key capacity bound~\cite{Pirandola.2017}, with its relative version scaled down by the detectors' efficiency.
    The Noisy \gls*{skc0} upper bound represents the realistic upper bound capacity for repeaterless quantum communications using detectors with the same efficiency and dark counts as those used in this experiment.
    Open symbols represent the state-of-the-art \gls*{skr} for field trial \gls*{qkd} tests using non-cryogenically cooled single photon detectors plotted at the fibre distance reported by the experiments (details in \supp).
    }
    \label{fig:SKR}
\end{figure*}

\section*{Discussion}
The obtained \gls*{skr} is depicted in \cref{fig:SKR} alongside fundamental bounds for point-to-point \gls*{qkd}. 
The \gls*{skr} surpasses the upper bound for the noisy \gls*{skc0} (yellow line in \cref{fig:SKR}), which can be interpreted as the upper bound for point-to-point \gls*{qkd} communications employing detectors with the same efficiency and dark counts as those used in this experiment.
This proves that our system effectively implements an untrusted quantum repeater over commercial telecom infrastructure for the first time.
Overcoming this bound, without using quantum memories or concatenated pure single photon sources, and at parity of detectors used, is only possible when optical coherence is maintained between the signals interfering at the central relay. 
Our result also closely approaches the relative \gls*{skc0} (dashed line), and simulations indicate that over a more symmetric link, our system would have overcome this bound at the same overall channel losses.
A detailed explanation of the plotted bounds and their analytical expression is provided in the \methods.

\Cref{fig:SKR} also compares our results with previous state-of-the-art \gls*{qkd} field trials using practical, non-cryogenic detectors.
Our experiment demonstrates the benefits of leveraging optical coherence in extending the maximum distance of quantum communications, effectively doubling the distance achieved in prior trials and increasing the tolerable loss budget by approximately three orders of magnitude.
The achieved \gls*{skr} also compares favourably, with the achieved rate sufficient to support low-rate one-time-pad encryption for critical data transfer or refresh AES-256 keys every few seconds, more than enough to support operation of commercial-off-the-shelf AES encryptors.

The system's performance could be further improved by cooling the \glspl*{apd} below \qty{-30}{\celsius}, which has been demonstrated feasible with thermoelectric coolers~\cite{Frohlich.2017}.
By operating at \qty{-60}{\celsius}, we estimate a reduction in the dark count rate by \qty{10}{\decibel} while retaining the same detection efficiency. 
This improvement would significantly extend the system’s range; with the \gls*{tfqkd} system’s \gls*{skr} scaling with the square root of channel loss, the operational range could increase by \qty{20}{\decibel}. 
This would raise the link loss budget from \qty{\AtoBLossLow}{\decibel} to \qty{\AtoBLossLowPlusTwentydB}{\decibel}, surpassing the state-of-the-art link loss of \qty{71.9}{\decibel} reported for cryogenically cooled point-to-point QKD systems~\cite{Boaron.2018}.

Our work represents the first demonstration of coherence-based quantum communications' compatibility with existing network infrastructure, and the first implementation of an effective quantum repeater over commercial networks.
We also achieved a new record distance for real-world \gls*{qkd} using non-cryogenically cooled detection technology and established a long-distance star-shaped \gls*{qkd} network with measurement-device-independent properties. 
Our findings confirm that the environmental conditions in operational telecommunication centres are comparable or even better than those simulated in labs, encouraging further commercialisation and prototyping of coherent quantum communication equipment.
Utilising practical solutions like frequency dissemination, regeneration, and off-band phase stabilisation, and equipment such as compact \gls*{ld}, \glspl*{im}, \glspl*{pm}, and \glspl*{apd}, this demonstration sets a foundation for future implementations of advanced quantum communication protocols requiring coherence maintenance~\cite{Gottesman.2012, Arrazola.2014}, such as DLCZ-based quantum repeaters~\cite{Duan.2001} and networks~\cite{Pompili.2021, Hermans.2022, Liu.01092023}.
This achievement, showcased over a highly asymmetric national-scale link, lays the groundwork for future high-performance, practical quantum communications.

\section*{Acknowledgments}
The authors acknowledge funding from the Ministry of Internal Affairs and Communications, Japan, via the project of ICT priority technology (JPMI00316) ``Research and Development for Construction of a Global Quantum Cryptography Network''.
The authors acknowledge funding from the European Union’s Horizon 2020 research and innovation programme under grant agreement No 857156 ``OPENQKD'', No  101072637 ``Quantum-Safe Internet (QSI)''.
The research leading to these results has received funding from the European Union’s Horizon Europe research and innovation programme under Grant Agreement No. 101100680 (GN5-1).
Co-funded by the European Union. Views and opinions expressed are however those of the author(s) only and do not necessarily reflect those of the European Union.
Neither the European Union nor the granting authority can be held responsible for them.

\section*{Authors contributions}
M.P. developed the experimental system, performed the experiment, provided the simulations, and analysed the data.
P.R., D.V., and G.R. provided access to the network infrastructure and supported the system installation.
Y.S.L., A.B., R.I.W., T.R., J.F.D., K.A.O. and S.J. supported the experimental work.
M.S.W. provided definitions for the secret capacity bounds.
M.P. and A.J.S. guided the work.
M.P. wrote the manuscript, with contributions from all the authors.

\section*{Data and materials availability}
All data are available in the manuscript or the supplementary materials.

\bibliography{TF_Field_trial_-_arXiv_-_Main}

\begin{thebibliography}{10}
\expandafter\ifx\csname url\endcsname\relax
  \def\url#1{\texttt{#1}}\fi
\expandafter\ifx\csname urlprefix\endcsname\relax\def\urlprefix{URL }\fi
\providecommand{\bibinfo}[2]{#2}
\providecommand{\eprint}[2][]{\url{#2}}

\bibitem{BB84}
\bibinfo{author}{Bennett, C.~H.} \& \bibinfo{author}{Brassard, G.}
\newblock \bibinfo{title}{Quantum cryptography: Public key distribution and coin tossing}.
\newblock \emph{\bibinfo{journal}{Theoretical Computer Science}} \textbf{\bibinfo{volume}{560}}, \bibinfo{pages}{7--11} (\bibinfo{year}{2014}).

\bibitem{Ekert.1991}
\bibinfo{author}{Ekert, A.~K.}
\newblock \bibinfo{title}{Quantum cryptography based on {B}ell's theorem}.
\newblock \emph{\bibinfo{journal}{Physical Review Letters}} \textbf{\bibinfo{volume}{67}}, \bibinfo{pages}{661--663} (\bibinfo{year}{1991}).

\bibitem{Xu.2020b}
\bibinfo{author}{Xu, F.}, \bibinfo{author}{Ma, X.}, \bibinfo{author}{Zhang, Q.}, \bibinfo{author}{Lo, H.-K.} \& \bibinfo{author}{Pan, J.-W.}
\newblock \bibinfo{title}{Secure quantum key distribution with realistic devices}.
\newblock \emph{\bibinfo{journal}{Reviews of Modern Physics}} \textbf{\bibinfo{volume}{92}}, \bibinfo{pages}{131} (\bibinfo{year}{2020}).

\bibitem{Pirandola.2020}
\bibinfo{author}{Pirandola, S.} \emph{et~al.}
\newblock \bibinfo{title}{Advances in quantum cryptography}.
\newblock \emph{\bibinfo{journal}{Advances in Optics and Photonics}} \textbf{\bibinfo{volume}{12}}, \bibinfo{pages}{1012} (\bibinfo{year}{2020}).

\bibitem{Buhrman.2001}
\bibinfo{author}{Buhrman, H.}, \bibinfo{author}{Cleve, R.}, \bibinfo{author}{Watrous, J.} \& \bibinfo{author}{de~Wolf, R.}
\newblock \bibinfo{title}{Quantum fingerprinting}.
\newblock \emph{\bibinfo{journal}{Physical Review Letters}} \textbf{\bibinfo{volume}{87}}, \bibinfo{pages}{167902} (\bibinfo{year}{2001}).

\bibitem{Buhrman.2010}
\bibinfo{author}{Buhrman, H.}, \bibinfo{author}{Cleve, R.}, \bibinfo{author}{Massar, S.} \& \bibinfo{author}{de~Wolf, R.}
\newblock \bibinfo{title}{Nonlocality and communication complexity}.
\newblock \emph{\bibinfo{journal}{Reviews of Modern Physics}} \textbf{\bibinfo{volume}{82}}, \bibinfo{pages}{665--698} (\bibinfo{year}{2010}).

\bibitem{Arrazola.2014}
\bibinfo{author}{Arrazola, J.~M.} \& \bibinfo{author}{L{\"u}tkenhaus, N.}
\newblock \bibinfo{title}{Quantum fingerprinting with coherent states and a constant mean number of photons}.
\newblock \emph{\bibinfo{journal}{Physical Review A}} \textbf{\bibinfo{volume}{89}}, \bibinfo{pages}{062305} (\bibinfo{year}{2014}).

\bibitem{Gottesman.2012}
\bibinfo{author}{Gottesman, D.}, \bibinfo{author}{Jennewein, T.} \& \bibinfo{author}{Croke, S.}
\newblock \bibinfo{title}{Longer-baseline telescopes using quantum repeaters}.
\newblock \emph{\bibinfo{journal}{Physical Review Letters}} \textbf{\bibinfo{volume}{109}}, \bibinfo{pages}{070503} (\bibinfo{year}{2012}).

\bibitem{Komar.2014}
\bibinfo{author}{K{\'o}m{\'a}r, P.} \emph{et~al.}
\newblock \bibinfo{title}{A quantum network of clocks}.
\newblock \emph{\bibinfo{journal}{Nature Physics}} \textbf{\bibinfo{volume}{10}}, \bibinfo{pages}{582--587} (\bibinfo{year}{2014}).

\bibitem{Degen.2017}
\bibinfo{author}{Degen, C.~L.}, \bibinfo{author}{Reinhard, F.} \& \bibinfo{author}{Cappellaro, P.}
\newblock \bibinfo{title}{Quantum sensing}.
\newblock \emph{\bibinfo{journal}{Reviews of Modern Physics}} \textbf{\bibinfo{volume}{89}}, \bibinfo{pages}{035002} (\bibinfo{year}{2017}).

\bibitem{Liu.2021a}
\bibinfo{author}{Liu, L.-Z.} \emph{et~al.}
\newblock \bibinfo{title}{Distributed quantum phase estimation with entangled photons}.
\newblock \emph{\bibinfo{journal}{Nature Photonics}} \textbf{\bibinfo{volume}{15}}, \bibinfo{pages}{137--142} (\bibinfo{year}{2021}).

\bibitem{Malia.2022}
\bibinfo{author}{Malia, B.~K.}, \bibinfo{author}{Wu, Y.}, \bibinfo{author}{Mart{\'i}nez-Rinc{\'o}n, J.} \& \bibinfo{author}{Kasevich, M.~A.}
\newblock \bibinfo{title}{Distributed quantum sensing with mode-entangled spin-squeezed atomic states}.
\newblock \emph{\bibinfo{journal}{Nature}} \textbf{\bibinfo{volume}{612}}, \bibinfo{pages}{661--665} (\bibinfo{year}{2022}).

\bibitem{Kimble.2008}
\bibinfo{author}{Kimble, H.~J.}
\newblock \bibinfo{title}{The quantum internet}.
\newblock \emph{\bibinfo{journal}{Nature}} \textbf{\bibinfo{volume}{453}}, \bibinfo{pages}{1023--1030} (\bibinfo{year}{2008}).

\bibitem{Tani.2005}
\bibinfo{author}{Tani, S.}, \bibinfo{author}{Kobayashi, H.} \& \bibinfo{author}{Matsumoto, K.}
\newblock \bibinfo{title}{Exact quantum algorithms for the leader election problem}.
\newblock In \bibinfo{editor}{Hutchison, D.} \emph{et~al.} (eds.) \emph{\bibinfo{booktitle}{STACS 2005}}, vol. \bibinfo{volume}{3404} of \emph{\bibinfo{series}{Lecture Notes in Computer Science}}, \bibinfo{pages}{581--592} (\bibinfo{publisher}{{Springer Berlin Heidelberg}}, \bibinfo{address}{Berlin, Heidelberg}, \bibinfo{year}{2005}).

\bibitem{Denchev.2008}
\bibinfo{author}{Denchev, V.~S.} \& \bibinfo{author}{Pandurangan, G.}
\newblock \bibinfo{title}{Distributed quantum computing}.
\newblock \emph{\bibinfo{journal}{ACM SIGACT News}} \textbf{\bibinfo{volume}{39}}, \bibinfo{pages}{77--95} (\bibinfo{year}{2008}).

\bibitem{Barz.2012}
\bibinfo{author}{Barz, S.} \emph{et~al.}
\newblock \bibinfo{title}{Demonstration of blind quantum computing}.
\newblock \emph{\bibinfo{journal}{Science}} \textbf{\bibinfo{volume}{335}}, \bibinfo{pages}{303--308} (\bibinfo{year}{2012}).

\bibitem{BenOr.05222005}
\bibinfo{author}{Ben-Or, M.} \& \bibinfo{author}{Hassidim, A.}
\newblock \bibinfo{title}{Fast quantum byzantine agreement}.
\newblock In \bibinfo{editor}{Gabow, H.} \& \bibinfo{editor}{Fagin, R.} (eds.) \emph{\bibinfo{booktitle}{Proceedings of the thirty-seventh annual ACM symposium on Theory of computing}}, \bibinfo{pages}{481--485} (\bibinfo{publisher}{ACM}, \bibinfo{address}{New York, NY, USA}, \bibinfo{year}{05222005}).

\bibitem{Pfaff.2014}
\bibinfo{author}{Pfaff, W.} \emph{et~al.}
\newblock \bibinfo{title}{Quantum information. unconditional quantum teleportation between distant solid-state quantum bits}.
\newblock \emph{\bibinfo{journal}{Science}} \textbf{\bibinfo{volume}{345}}, \bibinfo{pages}{532--535} (\bibinfo{year}{2014}).

\bibitem{Northup.2014}
\bibinfo{author}{Northup, T.~E.} \& \bibinfo{author}{Blatt, R.}
\newblock \bibinfo{title}{Quantum information transfer using photons}.
\newblock \emph{\bibinfo{journal}{Nature Photonics}} \textbf{\bibinfo{volume}{8}}, \bibinfo{pages}{356--363} (\bibinfo{year}{2014}).

\bibitem{Hensen.2015}
\bibinfo{author}{Hensen, B.} \emph{et~al.}
\newblock \bibinfo{title}{Loophole-free bell inequality violation using electron spins separated by 1.3 kilometres}.
\newblock \emph{\bibinfo{journal}{Nature}} \textbf{\bibinfo{volume}{526}}, \bibinfo{pages}{682--686} (\bibinfo{year}{2015}).

\bibitem{Roberts.2017c}
\bibinfo{author}{Roberts, G.~L.} \emph{et~al.}
\newblock \bibinfo{title}{Experimental measurement-device-independent quantum digital signatures}.
\newblock \emph{\bibinfo{journal}{Nature communications}} \textbf{\bibinfo{volume}{8}}, \bibinfo{pages}{1098} (\bibinfo{year}{2017}).

\bibitem{HerreroCollantes.2017}
\bibinfo{author}{Herrero-Collantes, M.} \& \bibinfo{author}{Garcia-Escartin, J.~C.}
\newblock \bibinfo{title}{Quantum random number generators}.
\newblock \emph{\bibinfo{journal}{Reviews of Modern Physics}} \textbf{\bibinfo{volume}{89}}, \bibinfo{pages}{90} (\bibinfo{year}{2017}).

\bibitem{Duan.2001}
\bibinfo{author}{Duan, L.~M.}, \bibinfo{author}{Lukin, M.~D.}, \bibinfo{author}{Cirac, J.~I.} \& \bibinfo{author}{Zoller, P.}
\newblock \bibinfo{title}{Long-distance quantum communication with atomic ensembles and linear optics}.
\newblock \emph{\bibinfo{journal}{Nature}} \textbf{\bibinfo{volume}{414}}, \bibinfo{pages}{413--418} (\bibinfo{year}{2001}).

\bibitem{Bennett.1993}
\bibinfo{author}{Bennett} \emph{et~al.}
\newblock \bibinfo{title}{Teleporting an unknown quantum state via dual classical and einstein-podolsky-rosen channels}.
\newblock \emph{\bibinfo{journal}{Physical Review Letters}} \textbf{\bibinfo{volume}{70}}, \bibinfo{pages}{1895--1899} (\bibinfo{year}{1993}).

\bibitem{Pompili.2021}
\bibinfo{author}{Pompili, M.} \emph{et~al.}
\newblock \bibinfo{title}{Realization of a multinode quantum network of remote solid-state qubits}.
\newblock \emph{\bibinfo{journal}{Science}} \textbf{\bibinfo{volume}{372}}, \bibinfo{pages}{259--264} (\bibinfo{year}{2021}).

\bibitem{Liu.01092023}
\bibinfo{author}{Liu, J.-L.} \emph{et~al.}
\newblock \bibinfo{title}{A multinode quantum network over a metropolitan area}.

\bibitem{Hermans.2022}
\bibinfo{author}{Hermans, S. L.~N.} \emph{et~al.}
\newblock \bibinfo{title}{Qubit teleportation between non-neighbouring nodes in a quantum network}.
\newblock \emph{\bibinfo{journal}{Nature}} \textbf{\bibinfo{volume}{605}}, \bibinfo{pages}{663--668} (\bibinfo{year}{2022}).

\bibitem{Lucamarini.2018}
\bibinfo{author}{Lucamarini, M.}, \bibinfo{author}{Yuan, Z.~L.}, \bibinfo{author}{Dynes, J.~F.} \& \bibinfo{author}{Shields, A.~J.}
\newblock \bibinfo{title}{Overcoming the rate-distance limit of quantum key distribution without quantum repeaters}.
\newblock \emph{\bibinfo{journal}{Nature}} \textbf{\bibinfo{volume}{557}}, \bibinfo{pages}{400--403} (\bibinfo{year}{2018}).

\bibitem{Wang.2018}
\bibinfo{author}{Wang, X.-B.}, \bibinfo{author}{Yu, Z.-W.} \& \bibinfo{author}{Hu, X.-L.}
\newblock \bibinfo{title}{Twin-field quantum key distribution with large misalignment error}.
\newblock \emph{\bibinfo{journal}{Physical Review A}} \textbf{\bibinfo{volume}{98}} (\bibinfo{year}{2018}).

\bibitem{Curty.2019}
\bibinfo{author}{Curty, M.}, \bibinfo{author}{Azuma, K.} \& \bibinfo{author}{Lo, H.-K.}
\newblock \bibinfo{title}{Simple security proof of twin-field type quantum key distribution protocol}.
\newblock \emph{\bibinfo{journal}{npj Quantum Information}} \textbf{\bibinfo{volume}{5}}, \bibinfo{pages}{64} (\bibinfo{year}{2019}).

\bibitem{Zeng.2022}
\bibinfo{author}{Zeng, P.}, \bibinfo{author}{Zhou, H.}, \bibinfo{author}{Wu, W.} \& \bibinfo{author}{Ma, X.}
\newblock \bibinfo{title}{Mode-pairing quantum key distribution}.
\newblock \emph{\bibinfo{journal}{Nature Communications}} \textbf{\bibinfo{volume}{13}}, \bibinfo{pages}{3903} (\bibinfo{year}{2022}).

\bibitem{Xie.2022}
\bibinfo{author}{Xie, Y.-M.} \emph{et~al.}
\newblock \bibinfo{title}{Breaking the rate-loss bound of quantum key distribution with asynchronous two-photon interference}.
\newblock \emph{\bibinfo{journal}{PRX Quantum}} \textbf{\bibinfo{volume}{3}} (\bibinfo{year}{2022}).

\bibitem{Avesani.2023}
\bibinfo{author}{Avesani, M.}
\newblock \bibinfo{title}{Long-range quantum cryptography gets simpler}.
\newblock \emph{\bibinfo{journal}{Physics}} \textbf{\bibinfo{volume}{16}}, \bibinfo{pages}{104} (\bibinfo{year}{2023}).

\bibitem{Pirandola.2017}
\bibinfo{author}{Pirandola, S.}, \bibinfo{author}{Laurenza, R.}, \bibinfo{author}{Ottaviani, C.} \& \bibinfo{author}{Banchi, L.}
\newblock \bibinfo{title}{Fundamental limits of repeaterless quantum communications}.
\newblock \emph{\bibinfo{journal}{Nature Communications}} \textbf{\bibinfo{volume}{8}}, \bibinfo{pages}{15043} (\bibinfo{year}{2017}).

\bibitem{Minder.2019}
\bibinfo{author}{Minder, M.} \emph{et~al.}
\newblock \bibinfo{title}{Experimental quantum key distribution beyond the repeaterless secret key capacity}.
\newblock \emph{\bibinfo{journal}{Nature Photonics}} \textbf{\bibinfo{volume}{13}}, \bibinfo{pages}{334--338} (\bibinfo{year}{2019}).

\bibitem{Pittaluga.2021}
\bibinfo{author}{Pittaluga, M.} \emph{et~al.}
\newblock \bibinfo{title}{600-km repeater-like quantum communications with dual-band stabilization}.
\newblock \emph{\bibinfo{journal}{Nature Photonics}} \textbf{\bibinfo{volume}{15}}, \bibinfo{pages}{530--535} (\bibinfo{year}{2021}).

\bibitem{Liu.2021}
\bibinfo{author}{Liu, H.} \emph{et~al.}
\newblock \bibinfo{title}{Field test of twin-field quantum key distribution through sending-or-not-sending over 428 km}.
\newblock \emph{\bibinfo{journal}{Physical Review Letters}} \textbf{\bibinfo{volume}{126}} (\bibinfo{year}{2021}).

\bibitem{Chen.2021}
\bibinfo{author}{Chen, J.-P.} \emph{et~al.}
\newblock \bibinfo{title}{Twin-field quantum key distribution over a 511 km optical fibre linking two distant metropolitan areas}.
\newblock \emph{\bibinfo{journal}{Nature Photonics}} \textbf{\bibinfo{volume}{299}}, \bibinfo{pages}{1476} (\bibinfo{year}{2021}).

\bibitem{Chen.2022}
\bibinfo{author}{Chen, J.-P.} \emph{et~al.}
\newblock \bibinfo{title}{Quantum key distribution over 658 km fiber with distributed vibration sensing}.
\newblock \emph{\bibinfo{journal}{Physical Review Letters}} \textbf{\bibinfo{volume}{128}}, \bibinfo{pages}{180502} (\bibinfo{year}{2022}).

\bibitem{Wang.2022}
\bibinfo{author}{Wang, S.} \emph{et~al.}
\newblock \bibinfo{title}{Twin-field quantum key distribution over 830-km fibre}.
\newblock \emph{\bibinfo{journal}{Nature Photonics}} \textbf{\bibinfo{volume}{16}}, \bibinfo{pages}{154--161} (\bibinfo{year}{2022}).

\bibitem{Clivati.2022}
\bibinfo{author}{Clivati, C.} \emph{et~al.}
\newblock \bibinfo{title}{Coherent phase transfer for real-world twin-field quantum key distribution}.
\newblock \emph{\bibinfo{journal}{Nature Communications}} \textbf{\bibinfo{volume}{13}}, \bibinfo{pages}{157} (\bibinfo{year}{2022}).

\bibitem{Liu.2023}
\bibinfo{author}{Liu, Y.} \emph{et~al.}
\newblock \bibinfo{title}{Experimental twin-field quantum key distribution over 1000 km fiber distance}.
\newblock \emph{\bibinfo{journal}{Physical Review Letters}} \textbf{\bibinfo{volume}{130}}, \bibinfo{pages}{210801} (\bibinfo{year}{2023}).

\bibitem{Zhou.2023}
\bibinfo{author}{Zhou, L.}, \bibinfo{author}{Lin, J.}, \bibinfo{author}{Jing, Y.} \& \bibinfo{author}{Yuan, Z.}
\newblock \bibinfo{title}{Twin-field quantum key distribution without optical frequency dissemination}.
\newblock \emph{\bibinfo{journal}{Nature Communications}} \textbf{\bibinfo{volume}{14}}, \bibinfo{pages}{928} (\bibinfo{year}{2023}).

\bibitem{Hadfield.2009}
\bibinfo{author}{Hadfield, R.~H.}
\newblock \bibinfo{title}{Single-photon detectors for optical quantum information applications}.
\newblock \emph{\bibinfo{journal}{Nature Photonics}} \textbf{\bibinfo{volume}{3}}, \bibinfo{pages}{696--705} (\bibinfo{year}{2009}).

\bibitem{Zhang.2015}
\bibinfo{author}{Zhang, J.}, \bibinfo{author}{Itzler, M.~A.}, \bibinfo{author}{Zbinden, H.} \& \bibinfo{author}{Pan, J.-W.}
\newblock \bibinfo{title}{Advances in ingaas/inp single-photon detector systems for quantum communication}.
\newblock \emph{\bibinfo{journal}{Light: Science {\&} Applications}} \textbf{\bibinfo{volume}{4}}, \bibinfo{pages}{e286--e286} (\bibinfo{year}{2015}).

\bibitem{Humer.2015}
\bibinfo{author}{Humer, G.} \emph{et~al.}
\newblock \bibinfo{title}{A simple and robust method for estimating afterpulsing in single photon detectors}.
\newblock \emph{\bibinfo{journal}{Journal of Lightwave Technology}} \textbf{\bibinfo{volume}{33}}, \bibinfo{pages}{3098--3107} (\bibinfo{year}{2015}).

\bibitem{Jiang.2020}
\bibinfo{author}{Jiang, C.}, \bibinfo{author}{Hu, X.-L.}, \bibinfo{author}{Xu, H.}, \bibinfo{author}{Yu, Z.-W.} \& \bibinfo{author}{Wang, X.-B.}
\newblock \bibinfo{title}{Zigzag approach to higher key rate of sending-or-not-sending twin field quantum key distribution with finite-key effects}.
\newblock \emph{\bibinfo{journal}{New Journal of Physics}} \textbf{\bibinfo{volume}{22}}, \bibinfo{pages}{053048} (\bibinfo{year}{2020}).

\bibitem{Lo.2012}
\bibinfo{author}{Lo, H.-K.}, \bibinfo{author}{Curty, M.} \& \bibinfo{author}{Qi, B.}
\newblock \bibinfo{title}{Measurement-device-independent quantum key distribution}.
\newblock \emph{\bibinfo{journal}{Physical Review Letters}} \textbf{\bibinfo{volume}{108}}, \bibinfo{pages}{130503} (\bibinfo{year}{2012}).

\bibitem{Braunstein.2012}
\bibinfo{author}{Braunstein, S.~L.} \& \bibinfo{author}{Pirandola, S.}
\newblock \bibinfo{title}{Side-channel-free quantum key distribution}.
\newblock \emph{\bibinfo{journal}{Physical Review Letters}} \textbf{\bibinfo{volume}{108}}, \bibinfo{pages}{130502} (\bibinfo{year}{2012}).

\bibitem{Ye.2003}
\bibinfo{author}{Ye, J.} \emph{et~al.}
\newblock \bibinfo{title}{Delivery of high-stability optical and microwave frequency standards over an optical fiber network}.
\newblock \emph{\bibinfo{journal}{Journal of the Optical Society of America B}} \textbf{\bibinfo{volume}{20}}, \bibinfo{pages}{1459} (\bibinfo{year}{2003}).

\bibitem{Comandar.2016c}
\bibinfo{author}{Comandar, L.~C.} \emph{et~al.}
\newblock \bibinfo{title}{Near perfect mode overlap between independently seeded, gain-switched lasers}.
\newblock \emph{\bibinfo{journal}{Optics express}} \textbf{\bibinfo{volume}{24}}, \bibinfo{pages}{17849--17859} (\bibinfo{year}{2016}).

\bibitem{Lo.2005}
\bibinfo{author}{Lo, H.-K.}, \bibinfo{author}{Ma, X.} \& \bibinfo{author}{Chen, K.}
\newblock \bibinfo{title}{Decoy state quantum key distribution}.
\newblock \emph{\bibinfo{journal}{Physical Review Letters}} \textbf{\bibinfo{volume}{94}}, \bibinfo{pages}{230504} (\bibinfo{year}{2005}).

\bibitem{Wang.2005}
\bibinfo{author}{Wang, X.-B.}
\newblock \bibinfo{title}{Beating the photon-number-splitting attack in practical quantum cryptography}.
\newblock \emph{\bibinfo{journal}{Physical Review Letters}} \textbf{\bibinfo{volume}{94}}, \bibinfo{pages}{230503} (\bibinfo{year}{2005}).

\bibitem{Yuan.2008}
\bibinfo{author}{Yuan, Z.~L.}, \bibinfo{author}{Dixon, A.~R.}, \bibinfo{author}{Dynes, J.~F.}, \bibinfo{author}{Sharpe, A.~W.} \& \bibinfo{author}{Shields, A.~J.}
\newblock \bibinfo{title}{Gigahertz quantum key distribution with ingaas avalanche photodiodes}.
\newblock \emph{\bibinfo{journal}{Applied Physics Letters}} \textbf{\bibinfo{volume}{92}} (\bibinfo{year}{2008}).

\bibitem{Zhang.2010}
\bibinfo{author}{Zhang, J.} \emph{et~al.}
\newblock \bibinfo{title}{2.23 ghz gating ingaas/inp single-photon avalanche diode for quantum key distribution}.
\newblock In \bibinfo{editor}{Itzler, M.~A.} \& \bibinfo{editor}{Campbell, J.~C.} (eds.) \emph{\bibinfo{booktitle}{Advanced Photon Counting Techniques IV}}, SPIE Proceedings, \bibinfo{pages}{76810Z} (\bibinfo{publisher}{SPIE}, \bibinfo{year}{2010}).

\bibitem{Hu.2019}
\bibinfo{author}{Hu, X.-L.}, \bibinfo{author}{Jiang, C.}, \bibinfo{author}{Yu, Z.-W.} \& \bibinfo{author}{Wang, X.-B.}
\newblock \bibinfo{title}{Sending-or-not-sending twin-field protocol for quantum key distribution with asymmetric source parameters}.
\newblock \emph{\bibinfo{journal}{Physical Review A}} \textbf{\bibinfo{volume}{100}} (\bibinfo{year}{2019}).

\bibitem{Frohlich.2017}
\bibinfo{author}{Fr{\"o}hlich, B.} \emph{et~al.}
\newblock \bibinfo{title}{Long-distance quantum key distribution secure against coherent attacks}.
\newblock \emph{\bibinfo{journal}{Optica}} \textbf{\bibinfo{volume}{4}}, \bibinfo{pages}{163} (\bibinfo{year}{2017}).

\bibitem{Boaron.2018}
\bibinfo{author}{Boaron, A.} \emph{et~al.}
\newblock \bibinfo{title}{Secure quantum key distribution over 421 km of optical fiber}.
\newblock \emph{\bibinfo{journal}{Physical Review Letters}} \textbf{\bibinfo{volume}{121}}, \bibinfo{pages}{190502} (\bibinfo{year}{2018}).

\bibitem{Xu.2020}
\bibinfo{author}{Xu, H.}, \bibinfo{author}{Yu, Z.-W.}, \bibinfo{author}{Jiang, C.}, \bibinfo{author}{Hu, X.-L.} \& \bibinfo{author}{Wang, X.-B.}
\newblock \bibinfo{title}{Sending-or-not-sending twin-field quantum key distribution: Breaking the direct transmission key rate}.
\newblock \emph{\bibinfo{journal}{Physical Review A}} \textbf{\bibinfo{volume}{101}} (\bibinfo{year}{2020}).

\bibitem{Pirandola.2019}
\bibinfo{author}{Pirandola, S.}
\newblock \bibinfo{title}{End-to-end capacities of a quantum communication network}.
\newblock \emph{\bibinfo{journal}{Communications Physics}} \textbf{\bibinfo{volume}{2}}, \bibinfo{pages}{1023} (\bibinfo{year}{2019}).

\bibitem{Takeoka.2014}
\bibinfo{author}{Takeoka, M.}, \bibinfo{author}{Guha, S.} \& \bibinfo{author}{Wilde, M.~M.}
\newblock \bibinfo{title}{Fundamental rate-loss tradeoff for optical quantum key distribution}.
\newblock \emph{\bibinfo{journal}{Nature Communications}} \textbf{\bibinfo{volume}{5}}, \bibinfo{pages}{5235} (\bibinfo{year}{2014}).

\bibitem{Rohde.2006}
\bibinfo{author}{Rohde, P.~P.} \& \bibinfo{author}{Ralph, T.~C.}
\newblock \bibinfo{title}{Modelling photo-detectors in quantum optics}.
\newblock \emph{\bibinfo{journal}{Journal of Modern Optics}} \textbf{\bibinfo{volume}{53}}, \bibinfo{pages}{1589--1603} (\bibinfo{year}{2006}).

\end{thebibliography}


\clearpage
\section{Methods}

\subsection{Overview of the \textit{service} and \textit{management} layers}
Simplified representations of the system and the signals' optical routing are reported in \cref{fig:sys_installation} and \cref{fig:sys_scheme} of the main text, respectively. 
A comprehensive schematic of the system is reported in the \supp.

The \textit{service layer}, the top layer in the functional diagram in \cref{fig:sys_installation} of the main text, comprises:
\begin{itemize}
    \item A VPN server in Alice, connected to the internet, providing remote access to all the equipment connected to this layer.
    \item Two optical fibres, one per duplex, connecting Alice to Charlie ($F_{C}^{AC}$) and Bob to Charlie ($F_{C}^{BC}$). 
    These are standard single-mode fibres, rented from the network provider, and injected with six optical signals at different wavelengths travelling in both directions (upstream and downstream).
    A full description of the signals transmitted over the $F_{C}$ fibres is reported in the \supp.
    \item Three optical MUX/DEMUX subsystems, one per node, injecting and extracting the optical signals into/from $F_{C}^{AC}$ and $F_{C}^{BC}$.
    \item Three Ethernet switches, one per node, endowed with RJ45 and SFP ports, are used to connect all the nodes' subsystems to the same local area network.
    \item A \qty{\SysRefClock}{\mega\hertz} RF reference clock in Charlie, generated locally by FPGA\textsuperscript{C} and distributed optically to FPGA\textsuperscript{A} and FPGA\textsuperscript{B}, where the signal are converted back to RF. 
    \item One seeding lasers subsystem in Charlie, comprising two CW lasers, $L_S$ and $L_Q$, emitting at $\lambda_S$=\qty{\LRefWavelength}{\nano\metre} and $\lambda_Q$=\qty{\LQuantWavelength}{\nano\metre}.
    Both lasers have an emission linewidth of a few hundred Hz.
    Outputs from $L_S$ and $L_Q$ are combined and sent to the transmitter nodes where they are used by the \textit{quantum layer} to encode the quantum communication protocol.
    \item A \qty{120}{\kilo\metre} long fibre, followed by an \gls*{edfa} (shown in the detailed schematics in the \supp), delay and amplify the $\lambda_S$ and $\lambda_Q$ signals for Bob before these are injected into $F_{C}^{BC}$.
    This delay ensures approximate matching of the overall optical path of Alice's and Bob's $\lambda_S$ and $\lambda_Q$ signals within the coherence length of their respective laser sources.
    This approach enables the execution of a coherence-based quantum communication protocol over a highly asymmetric channel and it could be removed using master laser sources with suitably narrow emission linewidth.    
\end{itemize}

The \textit{management layer}, the middle layer in the functional diagram shown in \cref{fig:sys_installation}, consists of the following components:
\begin{itemize}
    \item Computer servers (Server\textsuperscript{A,B,C}): Each node has a dedicated server connected to the same network by the \textit{service layer}.
    These servers manage all the equipment at their respective nodes.
    \item FPGAs (FPGA\textsuperscript{A,B,C}): One FPGA per node ensures synchronization of the reference clock and absolute time among the nodes.
    The timing signals distributed by the service layer facilitate this synchronization.
    Notably, the jitter between the three clocks of the nodes was tested in the lab and found to be smaller than \qty{8}{\pico\second}.
    \item Two \glspl*{awg}: Located at the transmitter nodes, these \glspl*{awg} are locked to the reference clock provided by FPGAs A and B.
    They generate \unit{\giga\hertz} driving signals used by the \textit{quantum layer} to encode the quantum communication protocol.
\end{itemize}

\subsection{Overview of the \textit{quantum} layer}
The \textit{quantum layer}, shown at the bottom of the functional diagram in \cref{fig:sys_installation}, comprises five subsystems: the Quantum MUX/DEMUX unit and the Quantum Encoder at the transmitters, and the Quantum Receiver, four \glspl*{apd} and a time tagger at the receiver node.

Each Quantum DEMUX/MUX unit implements the following operations:
\begin{itemize}
    \item It amplifies the $\lambda_S$ and $\lambda_Q$ signals through an \gls*{edfa} optical amplifier.
    \item It separates $\lambda_S$ from $\lambda_Q$ with a frequency demultiplexer, 
    and passes $\lambda_Q$ to the Quantum Encoder. 
    \item It manipulates the intensity and polarisation of the $\lambda_S$ signal according to the requirement for the stabilisation of the optical channel. The manipulated signal is labelled $\lambda'_S$.
    \item It multiplexes back together the tuned $\lambda'_S$ signal with the encoded $\lambda'_Q$ coming from the Quantum Encoder.
    \item It finally sends $\lambda'_S$ and $\lambda'_Q$ to Charlie by injecting them into the $F_{Q}$ fibres connecting the transmitters to the receiver node.
\end{itemize}

Each Quantum Encoder transforms the received $\lambda_Q$ \gls*{cw} signal into a $\lambda'_Q$ signal, modulated according to the specifications of the SNS TF-QKD protocol.
Within the Encoder, $\lambda_Q$ is injected into a laser diode tuned for optimal \gls*{oil} with this signal.
The laser diode's output undergoes processing by three \glspl*{im} arranged in series, which carve the \gls*{cw} signal into \qty{1}{\giga\hertz}, \qty{\PulsesDuration}{\pico\second}-long optical pulses modulated at four different intensities.
Subsequently, two \glspl*{pm} placed in series modulate the phase of these pulses with arbitrary phase within the $\left[0,2\pi\right)$ interval.
These five high-speed electro-optics modulators are followed by a \gls*{voa}, setting the average photon flux to the correct level, and an \gls*{epc} for precompensating the polarisation rotation introduced by the quantum channel.

When Alice's and Bob's modulated $\lambda'_S$ and $\lambda'_Q$ signals enter the quantum receiver, they each encounter a \gls*{pbs} which separates the light into horizontal ($\lambda'^{\paral}$) and vertical ($\lambda'^{\perp}$) polarisation components.
The four vertical polarisation components $\lambda'^{A,\perp}_{S,Q}$ and $\lambda'^{B,\perp}_{S,Q}$ are monitored in turn by APD\textsubscript{4} which provides the information to maximise the intensity of the complementary horizontal polarisation components.
The horizontal components $\lambda'^{B,\paral}_{S,Q}$ from Bob pass then through a phase modulator, used to stabilise the optical interference.
After this, $\lambda'^{A,\paral}_{S,Q}$ and $\lambda'^{B,\paral}_{S,Q}$ interfere on a polarisation maintaining \gls*{bs}.
The two outputs of the interference between $\lambda'^{A,\paral}_{Q}$ and $\lambda'^{B,\paral}_{Q}$ are monitored by APD\textsubscript{1} and APD\textsubscript{2}.
Clicks collected by these detectors are analysed by a time tagger which provides the time-of-arrival information for each photon detection that is necessary for the protocol execution.
The outputs of the interference between $\lambda'^{A,\paral}_{S}$ and $\lambda'^{B,\paral}_{S}$ can be monitored in turn by APD\textsubscript{3}.
The clicks recorded by this detector are used to implement the fast feedback of the off-band phase stabilisation.

\subsection{Off-band phase stabilisation}
In this technique, two optical signals close in frequency are multiplexed and transmitted together and thus pick up an almost identical phase noise introduced by the transmission channels.
One of the two wavelengths, $\lambda_S$, acts as a support for the channel stabilisation.
$\lambda_S$ is kept as a \gls*{cw} signal and is modulated only in intensity and polarisation to optimise the optical interference in Charlie between the $\lambda'^A_S$ and $\lambda'^B_S$ sent by Alice and Bob respectively.
By actively stabilising at Charlie the interference between $\lambda'^A_S$ and $\lambda'^B_S$ through a fast, \textit{coarse}, feedback loop driving a phase modulator, all the noise introduced by the channels is corrected at $\lambda_S$.
The same phase correction is also applied to $\lambda_Q$, the wavelength used for the encoding of the quantum protocol, and this induces a reduction of more than three orders of magnitude to the phase noise picked up by $\lambda_Q$ as well.
The signal at $\lambda_Q$ is fully stabilised by a second slower, \textit{fine}, phase feedback acting on $\lambda_Q$ alone and implemented through a fibre stretcher.
This technique is instrumental to remove the phase noise introduced by the transmission channels, thus permitting the execution of a phase-based quantum communication protocol.

\subsection{The AOPP SNS TF-QKD protocol}
In this experiment, we implemented the 4-intensity SNS-TF-QKD quantum communication protocol~\cite{Wang.2018} over a highly asymmetric link~\cite{Hu.2019}.
Our implementation incorporated the \gls*{aopp} technique~\cite{Xu.2020, Jiang.2020} during post-processing to enhance the signal-to-noise ratio.
We also accounted for finite-size effects in our analysis.

The protocol involves repeating the following steps $N_{tot}$ times, until sufficient raw key material is generated for post-processing.

\textbf{Preparation of Weak Coherent States:} 
    Alice and Bob independently prepare weak \gls*{wcp} of the form $\ket{\sqrt{\mu}e^{i\theta}}$, where the pulses intensity $\left|\mu\right|^2$ and phase $\theta$ are determined by the specific encoding chosen by the users.

\textbf{Selection of Code or Test Basis:} 
    Alice (Bob) first selects the code $\mathbb{Z}$ or test - or decoy - $\mathbb{X}$ basis with probability $P_{\mathbb{Z}_A}$ ($P_{\mathbb{Z}_B}$) and $P_{\mathbb{X}_A}$ ($P_{\mathbb{X}_B}$)
    respectively, where $P_{\mathbb{Z}} + P_{\mathbb{X}} = 1 $.

\textbf{Pulses Encoding}: 
    When the code basis $P_{\mathbb{Z}}$ is chosen, Alice (Bob) encodes a bit 1 (0) by preparing a phase-randomized \gls*{wcp} with intensity $\left|\mu\right|^2=s_A$ ($s_B$) with probability $\epsilon_A$ ($\epsilon_B$), and encodes a bit 0 (1) by preparing a vacuum state $\left|\mu\right|^2=n_A$ ($n_B$) with probability $1-\epsilon_A$ ($1-\epsilon_B$).
    When the $P_{\mathbb{Z}}$ basis is chosen, Alice (Bob) randomly selects a flux value $\left|\mu\right|^2 = \left\{u_A, v_A, w_A\right\}$ ($\left\{u_B, v_B, w_B\right\}$) with conditional probability $\left\{ P_{u|\mathbb{X}_A}, P_{v|\mathbb{X}_A}, P_{w|\mathbb{X}_A} \right\}$ ($\left\{ P_{u|\mathbb{X}_B}, P_{v|\mathbb{X}_B}, P_{w|\mathbb{X}_B} \right\}$), where $P_{u|\mathbb{X}} + P_{v|\mathbb{X}} + P_{w|\mathbb{X}} = 1$, and a random global phase value $\varphi \in \left[ 0, 2\pi\right)$.
    Due to the asymmetry of the link, the encoding parameters that maximise the \gls*{skr} in our experimental condition are different.
    In this scenario, the following relation between the encoding parameters must be met to ensure the protocol security as shown in \cite{Hu.2019}:
    
    \begin{align}
        \frac{v_A}{v_B} = \frac{\epsilon_A \left(1-\epsilon_B\right) \, s_A \, e^{-s_A}}{\epsilon_B \left(1-\epsilon_A\right) \, s_B \, e^{-s_B}}
        \label{eq:asymm_security_condition}
    \end{align}

\textbf{Pulses Transmission and Detection Announcement}: 
    After the preparation stage, Alice and Bob send their encoded \gls*{wcp} to Charlie, who interferes them on a 50:50 \gls*{bs} followed by two single-photon detectors monitoring its two outputs.
    When only a single detector clicks per interference, Charlie broadcasts which one clicked over a public communication channel.
    At this point, Alice and Bob check what basis they used for each successful detection announced by Charlie.
    They label events where they both encoded on the $\mathbb{X}$ basis as $\mathbb{X}$-windows, and events where they both encoded on the $\mathbb{Z}$ as $\mathbb{Z}$-windows.

\textbf{Parameters Estimation and SKR Calculation}:
    Events in the $\mathbb{X}$-windows are used for the security analysis.
    The phase error rate can be extracted from these events by post-selecting instances where Alice and Bob encoded the same decoy state (e.g., $u_A$ for Alice and $u_B$ for Bob, or $v_A$ and $v_B$), with the \gls*{wcp} satisfying the following phase matching condition:
    \begin{align}
        \left| \theta_A - \theta_B \right| \leq \frac{2\pi}{M} \qquad \text{or} \qquad \left| \theta_A - \theta_B - \pi \right| \leq \frac{2\pi}{M}
    \end{align}

    From their respective encoding associated with $\mathbb{Z}$-windows, Alice and Bob can each extract a raw key of length $n_t$.
    The discrepancies between these two strings constitute the bit-flip error rate $E_Z$.
    The \gls*{aopp} method is a post-processing technique requiring two-way classical communication between Alice and Bob that reduces $E_Z$ at the cost of reducing the length of $n_t$.
    In \gls*{aopp}, Bob pairs up all the bits 0 in his raw strings with bits 1.
    He then communicates to Alice the positions of the pairs.
    Alice then pairs the bits in the same manner as Bob and checks their parity. 
    She then discards all the resulting pairs with even parity, and communicates the position of the surviving pairs to Bob.
    They then discard the second bit of the surviving pairs from their strings and each creates a second, shorter, string $n'_t$ with a considerably smaller bit-flip error rate $E'_Z$, which they will use to extract the final key.
    The resulting shorter string $n'_t$ with reduced error $E'_Z$ enhances secret key generation, enabling longer-distance communication with higher \gls*{skr}.

    The decoy-state method enables precise estimation of the number of single photon quantities needed for the \gls*{skr} calculation through analysis of the events recorded in the $\mathbb{X}$-windows.
    In the following, we denote the counting rate associated with the pulses of intensity $\mu_A$ ($\mu_A = u_A, v_A, w_A$) for Alice and $\mu_B$ ($\mu_B = u_B, v_B, w_B$) for Bob as $S_{\mu_A,\mu_B} = n_{\mu_A,\mu_B} / N_{\mu_A,\mu_B}$, where $N_{\mu_A,\mu_B}$ is the number of pulse pairs of source $ \mu_A, \mu_b $ sent out in the whole protocol, and $n_{\mu_A,\mu_B}$ the total number of one-detector heralded events of source $\mu_A, \mu_b$.
    $S_{\mu_A,\mu_B}$ can be measured experimentally.
    Using the recorded countrate, and the information on the photon fluxes, it is possible to estimate the lower bounds of the expected values of the counting rate of states $\ketbra{01}{01}$ and $\ketbra{10}{10}$, which are
    \begin{align}
        \resizebox{\columnwidth}{!}{$
            \avg{\underline{s_{01}}} = \frac{
                                                u_B^2 e^{v_B} \avg{\underline{S}_{w_A,v_B}} 
                                                - v_B^2 e^{u_B} \avg{\overline{S}_{w_A,u_B}} 
                                                - (u_B^2-v_B^2) \avg{\overline{S}_{w_A,w_B}} 
                                            }
                                            {u_B v_B (u_B - v_B)}
        $}
    \end{align}

    \begin{align}
        \resizebox{\columnwidth}{!}{$
            \avg{\underline{s_{10}}} = \frac{
                                                u_A^2 e^{v_A} \avg{\underline{S}_{v_A,w_B}} 
                                                -v_A^2 e^{u_A} \avg{\overline{S}_{u_A,w_B}}  
                                                -(u_A^2 - v_A^2) \avg{\overline{S}_{w_A,w_B}} 
            }{u_A v_A (u_A - v_A)}
        $}
    \end{align}
    where $\avg{\underline{\cdot}}$ and $\avg{\overline{\cdot}}$ represent the lower and the upper bound of the corresponding expected values, with a composable definition of security and the Chernoff bound being applied. 
    The lower bound of the expected value of the counting rate of untagged (i.e., single) photons is then:
    \begin{align}
        \avg{\underline{s_1}} = \frac{v_A}{v_A+v_B} \avg{\underline{s_{10}}} 
                                +\frac{v_B}{v_A+v_B} \avg{\underline{s_{01}}}
    \end{align}
    
    The lower bounds of the expected value of the number of untagged 0-bits, $\avg{\underline{n_{01}}}$, and untagged 1-bits, $\avg{\underline{n_{01}}}$, can be calculated as:
    \begin{align}
        \avg{\underline{n_{01}}} = N_{tot} P_{\mathbb{Z}_A} P_{\mathbb{Z}_B} \epsilon_A (1-\epsilon_B) s_A e^{-s_A} \avg{\underline{s_{10}}}
    \end{align}
    
    \begin{align}
        \avg{\underline{n_{01}}} = N_{tot} P_{\mathbb{Z}_A} P_{\mathbb{Z}_B} \epsilon_B (1-\epsilon_A) s_B e^{-s_B} \avg{\underline{s_{01}}}
    \end{align}
    The expected value of the lower bounds of the number of untagged bits is:
    \begin{align}
        \avg{\underline{n_{1}}} = \avg{\underline{n_{01}}} + \avg{\underline{n_{10}}} 
    \end{align}

    The phase-flip error rate of untagged bits in $\mathbb{Z}$-windows, $\avg{\overline{e_{1}^{ph}}}$, can be calculated from the bit-flip error rate of untagged bits in $\mathbb{X}$ windows as
    \begin{align}
         \avg{\overline{e_{1}^{ph}}} = \frac{ 
                                                \avg{\overline{T_{X_1}}} - e^{-v_A-v_B} \avg{\overline{S}_{w_A,w_B}} / 2
                                            }
                                            {
                                                e^{-v_A-v_B} (v_A+v_B) \avg{\underline{s_1}}
                                            }
    \end{align}
    where $\avg{\overline{T_{X_1}}}$ is the ratio between the total number of pulses sent out with intensities $v_A$ and $v_B$, and the number of corresponding error events. 
    
The \gls*{skr} of finite size SNS-TF-QKD with \gls*{aopp} is calculated as:
\begin{align}
    R = \frac{2}{N_{tot}} \left\{ n'_1 \left[ 1 - h \left( e'^{ph}_1 \right) \right] - f_{EC} \, n'_t \, h \left( E'_Z \right) -  \Delta_{FS} \right\} 
\end{align}
where 
$h \left( x \right) = -x \log_2 x - \left( 1 - x \right) \log_2 \left( 1 - x \right)$ is the Shannon entropy,
$f_{EC} = 1.05$ the error correction efficiency factor, 
and $\Delta_{FS} = \log_2 \frac{2}{\epsilon_{cor}} + 2 \log_2 \frac{1}{\sqrt{2} \epsilon_{PA} \hat{\epsilon}}$ is the finite-size correction term, 
with $\epsilon_{cor} = 10^{-10}$ the error correction failure probability, 
$\epsilon_{PA} = 10^{-10}$ the privacy amplification failure probability, 
and $\hat{\epsilon} = 10^{-10}$ the smoothing parameter for the min- and max-entropy chain rules.
$n'_1$ is the number of the untagged bits after AOPP, and $e'^{ph}_1$ their phase-flip error rate.

Further details regarding the \gls*{skr} calculation process can be found in ref.~\cite{Jiang.2020}, which was closely followed in this study's computations, mirroring methodologies employed in previous experimental works \cite{Pittaluga.2021, Liu.2021, Chen.2022, Zhou.2023}. 

\subsection{Secret key capacity bounds}
Our quantum communications system consists of a simple quantum network composed of three nodes in a chain: two end users, Alice (A) and Bob (B), and a middle user, Charlie (C). 
The network spans between the cities of Frankfurt and Kehl, with the central relay positioned in Kirchfeld at an asymmetric distance, approximately three-fifths of the total span.

The two-way assisted capacity $\mathcal{C}$ determines the optimal end-to-end rate of quantum communications achievable without restrictions on local operations and classical communication, which can be unlimited and two-way~\cite{Pirandola.2017}.
Our goal was to implement \gls*{qkd}, then we note that the \gls*{skc0} is equal to $\mathcal{C}$~\cite{Pirandola.2017}.
The ultimate end-to-end capacity and secret key capacity of a simple asymmetric one-repeater network for pure-loss channels is given by~\cite{Pirandola.2019}
\begin{align}
    \text{SKC}_1(\eta_A,\eta_B) = -\log_2{[1-\text{min}(\eta_A,\eta_B)]}\text{ ebits/use},
\end{align}
where $\eta_A$ and $\eta_B$ are the transmissivities of the lossy channels separating Alice and Bob, respectively, from Charlie.
The subscript in the notation \acrshort*{skc1} refers to the number of quantum repeaters.
The channel transmissivities can be asymmetric, i.e., when $\eta_A\neq\eta_B$.
These capacities are plotted in \cref{fig:SKR}.
We plot \acrshort*{skc1} for a centred Charlie (green) and \acrshort*{skc1} asymmetric for an asymmetric configuration as in our system (light green).

At low transmissivities, the single-repeater capacity scales like the worst transmissivity of the network, i.e.,  $\text{SKC}_1(\eta_A,\eta_B) \approx 1.44\; \text{min}(\eta_A,\eta_B)$.
While the highest rate of any implementation of \gls*{tfqkd} cannot overcome the single-repeater capacity, \gls*{tfqkd} can achieve the optimal rate-distance scaling, up to a constant factor.
This optimal scaling can be observed in \cref{fig:SKR} of the main text for the simulation of our \gls*{tfqkd} implementation (blue).

In contrast, the rate of any point-to-point quantum communication protocol scales like the total transmissivity with distance~\cite{Takeoka.2014}.
The fundamental rate-distance limit without a repeater is given by the repeaterless capacity~\cite{Pirandola.2017}
\begin{align}
    \text{SKC}_0(\eta_\text{channel}) = -\log_2{(1-\eta_\text{channel})}\text{ ebits/use},
\end{align}
where $\eta_\text{channel} = \eta_A \eta_B$ is the total channel transmissivity of the lossy channel separating Alice and Bob.
This two-way assisted capacity represents the ultimate rate that is reachable without a quantum repeater and is plotted in \cref{fig:SKR} (red).
The rate of our system (star marker) closely approaches the $\text{SKC}_0(\eta_\text{channel})$ without overcoming it.
However, to assess the advantages of adopting a coherence-based \gls*{qkd} protocol over other point-to-point repeaterless \gls*{qkd} implementations, we can benchmark our results against the upper bound of the maximum rate achievable by a point-to-point system using detectors with similar properties (efficiency and dark counts) to those used in this experiment.

The upper bound on the noisy repeaterless capacity for a channel transmissivity $\eta_\text{channel} = \eta_A \eta_B$, detector efficiency $\epsilon$, and dark-count rate $d_c$, is given by the upper bound on the thermal-loss channel capacity $\text{SKC}^\text{noisy}_0(\eta_\text{total}, \bar{n})$ characterised by total transmissivity $\eta_\text{total} = \eta_\text{channel} \epsilon$ and mean photon number $\bar{n} \approx d_c/(1-\eta_\text{total})$, that is~\cite{Pirandola.2017}
\begin{align}
    \text{SKC}^\text{noisy}_0(\eta_\text{total}, \bar{n}) \leq \text{UB}(\eta_\text{total}, \bar{n})\text{ ebits/use},\label{eq:UB}
\end{align}
where $\text{UB}(\eta_\text{total}, \bar{n}) = -\log_2[(1-\eta_\text{total}) \eta^{\bar{n}}]- h(\bar{n})$ for $\bar{n} < \eta_\text{total}/(1-\eta_\text{total})$,
and $h(x) \equiv (x+1) \log_2(x+1) - x \log_2(x) $.
We plot this noisy SKC$_0$ upper bound in \cref{fig:SKR} of the main text (solid yellow). 
A repeater is required to overcome this upper bound assuming Alice and Bob use non-ideal detectors with untrusted efficiency $\epsilon$ and dark-count rate $d_c$.
\Cref{fig:SKR} of the main text clearly shows that our system produces a secret key rate higher than the noisy SKC$_0$ upper bound, meaning that no point-to-point protocol using similar detectors can achieve our rate.
We derive this result in the following paragraphs.

We first model 
detectors with non-unit efficiency $\epsilon$ by including additional loss before the detector with transmissivity $\epsilon$~\cite{Rohde.2006}, then the capacity adjusted for inefficient detectors becomes
\begin{align}
    \text{SKC}_0(\eta_\text{total}) = -\log_2{(1-\eta_\text{total})}\text{ ebits/use},
\end{align}
where $\eta_\text{total} = \eta_\text{channel} \; \epsilon$ (pink dashed line in \cref{fig:SKR}).

Still, the capacity assumes no additional sources of noise.
The primary source of noise other than pure loss comes from detector dark counts which are false detection events at the detectors.
We can simultaneously model dark counts and non-ideal efficiency by including a Gaussian thermal-loss channel before each detector~\cite{Rohde.2006}.
That is, we assume that the dark counts are originating in the channel, and for any channel transmissivity $\eta_\text{channel}$, detector efficiency $\epsilon$ and dark-count rate $d_c$, there is an equivalent thermal-loss channel characterised by the total transmissivity $\eta_\text{total}$ and mean photon number $\bar{n}$.
While the exact achievable capacity of the thermal-loss channel is unknown, an upper bound is known~\cite{Pirandola.2017} and was given in~\cref{eq:UB}.

The thermal-loss channel with transmissivity $\eta_\text{total}$ and mean-photon number $\bar{n}$ is equivalent to interfering a thermal state with mean photon number $\bar{n}$ with the input state on a beamsplitter of transmissivity $\eta_\text{total}$, where the thermal state is defined
\begin{align}
    \rho_{\bar{n}} &= \sum_{n=0}^\infty \frac{\bar{n}^n}{(\bar{n}+1)^{n+1}} \ketbra{n}{n}.
\end{align} 
Then, the dark count rate is defined as the probability the detector registers ``ON'' when no photons are input onto it, i.e., when the input state is vacuum.

Since our detectors have low dark count rates ($d_c < 10^{-6}$), this translates to a small value of $\bar{n}$. To first order in $\bar{n}$ the thermal state is
\begin{align}
    \rho_{\bar{n}} &\approx \frac{1}{\bar{n}+1} \ketbra{0}{0} +  \frac{\bar{n}}{(\bar{n}+1)^{2}} \ketbra{1}{1} \approx \ketbra{0}{0} + \bar{n} \ketbra{1}{1}.
\end{align} 
That is, the photon is prepared with probability $\frac{\bar{n}}{(\bar{n}+1)^{2}} \approx \bar{n}$ and reflected off the beamsplitter with probability $1-\eta_\text{total}$ giving probability to be detected ``ON'' at the detector
\begin{align}
    P(\text{``ON''}) &= d_c \approx (1-\eta_\text{total}) \bar{n}.
\end{align}
Thus, to first order in $\bar{n}$ we find that 
\begin{align}
    \bar{n} \approx \frac{d_c}{1-\eta_\text{total}}.
\end{align}
This approximation is sufficient for presenting results. 
Numerically computing $\bar{n}$ to higher order in photon number has no visually noticeable change to the upper bound plotted in \cref{fig:SKR} of the main text.


\normalsize 

\makeatletter
\renewcommand \thesection{S\@arabic\c@section}
\renewcommand \thetable{S\@arabic\c@table}
\renewcommand \thefigure{S\@arabic\c@figure}
\makeatother

\setcounter{figure}{0}
\setcounter{table}{0}

\clearpage
\onecolumngrid
\section{Supplementary Materials}
\subsection{Characterisation of the Fibre Links}\label{supp:fibre_characterisation}

The experiment's location was selected from fibre links available to G\'EANT, considering suitability for the \gls*{tfqkd} protocol in terms of link asymmetry and loss budget. 
Nodes in the network are interconnected via fibre duplex cables.
Fibre duplexes are a typical resource in current telecommunication infrastructure.
Normally, they are used to implement full-duplex communication with each fibre carrying a unidirectional stream of data: one transmitting data (TX) and the other receiving data (RX).

In this experiment, each duplex includes one fibre, labelled \textit{classical fibre} ($F_C$), carrying a bidirectional data stream.
The other fibre, named \textit{quantum fibre} ($F_Q$), transmits unidirectional encoded quantum ($\lambda'_Q$) and supports ($\lambda'_S$) signals from the transmitters to the central node. 
A comprehensive description of all the signals carried by the \textit{classical fibres} is provided in \cref{supp:sec:optical_signals_description}.

All fibres were characterised before the experiment via \gls*{otdr} and channel loss measurements. 
\Cref{supp:tab:Fibre_properties} outlines the losses and lengths of the deployed fibres.
For each duplex, the fibre with the lowest losses was designated as the \textit{quantum fibre}, aiming to minimise the overall loss in the quantum channel and enhance quantum communication performance.

Each fibre, $F^{AC}_Q$ and $F^{AC}_C$, linking Frankfurt to Kirchfeld (Alice's and Charlie's locations, respectively), comprises two fibre spans joined in a colocation centre in Lampertheim, positioned approximately midway.

\begin{table}[!ht]
    \centering
        \begin{tabular}{lcclc}
            \toprule
                                                                            & Label         & \multirow{2}{*}{\shortstack[c]{Attenuation\\(dB)}}    & Purpose                   & \multirow{2}{*}{\shortstack[c]{Length\\(km)}}     \\
                                                                            &               &                                                       &                           &                                                   \\
            \midrule
                \multirow{2}{*}{Frenkfurt $\leftrightarrow$ Kirchfeld}      & $F^{AC}_Q$    & \AtoCLossLow                                          & Quantum Protocol          & \multirow{2}{*}{156.7}                            \\
                                                                            & $F^{AC}_C$    & \AtoCLossHigh                                         & Classical Communications  &                                                   \\
            \midrule
                \multirow{2}{*}{Kirchfeld $\leftrightarrow$ Kehl}           & $F^{BC}_Q$    & \BtoCLossLow                                          & Quantum Protocol          & \multirow{2}{*}{97.2}                             \\
                                                                            & $F^{BC}_C$    & \BtoCLossHigh                                         & Classical Communications  &                                                   \\
            \bottomrule
        \end{tabular}
    \caption{
    Properties of the deployed fibres used in the experiment.
    }\label{supp:tab:Fibre_properties}
\end{table}

\subsection{Signals Transmitted Over the Classical Fibres}\label{supp:sec:optical_signals_description}

The \textit{classical communication layer} includes two optical fibres connecting Charlie to Alice ($F_{C}^{AC}$) and to Bob ($F_{C}^{BC}$).
These fibres carry several optical signals at different wavelengths travelling in two directions, upstream and downstream.
The routing of the optical signals in the system is optimised for long-distance communication and is represented in \cref{fig:sys_scheme} of the main text, and in more detail in \cref{fig:supp:full_sys}.
Upon reaching the designated receiving node, all transmitted signals are amplified using \glspl*{edfa}, and signals at different optical frequencies are separated using \glspl*{dwdm}.
The optical wavelength and purpose of all the signals transmitted over the \textit{classical fibres} are summarised in \cref{supp:tab:optical_signals_description}. 

\begin{table*}[!ht]
    \centering
    \resizebox{\textwidth}{!}{
        \begin{tabular}{lllll}
            \toprule
            \multirow{2}{3em}{\textbf{ITU CH}} & \multirow{2}{6em}{\textbf{Wavelength (nm)}}         & \textbf{Purpose}          & \multirow{2}{6em}{\textbf{Signal direction}}        & \textbf{Description}               \\
                            &                   &                               &                               &                           \\
            \midrule
            CH32            & 1551.72           & Protocol reference            & C $\rightarrow$ A,B           & CW optical reference for the quantum layer - for $\lambda_Q$                                              \\
            CH34            & 1550.12           & Protocol signal               & C $\rightarrow$ A,B           & CW optical reference for the quantum layer - for $\lambda_S$                                              \\
            CH55            & 1533.47           & FPGAs direct link (dwnlink)   & C $\rightarrow$ A,B           & \multirow{2}{30em}{Optical signal from FPGA\textsubscript{C} to FPGA\textsubscript{A,B} for direct FPGA communication and time synchronisation}  \\
                            &                   &                               &                               &                                                                                                           \\
            CH56            & 1532.68           & Ethernet downlink             & C $\rightarrow$ A,B           & Optical signal from Charlie's Ethernet switch to Alice and Bob                                            \\
            CH57            & 1531.90           & FPGAs direct link (uplink)    & A,B $\rightarrow$ C           & \multirow{2}{30em}{Optical signal from FPGA\textsubscript{A,B} to FPGA\textsubscript{C} for direct FPGA communication and time synchronisation} \\
                            &                   &                               &                               &                                                                                                           \\
            CH58            & 1531.12           & Ethernet uplink               & A,B $\rightarrow$ C           & Optical signals from Alice and Bob's Ethernet switches to Charlie                                         \\
            \bottomrule
        \end{tabular}
    }
    \caption{
    Description of the optical signals transmitted over the classical fibres.
    }\label{supp:tab:optical_signals_description}
\end{table*}

\subsection{Comparison of the Phase Noise between of installed fibres}\label{supp:fibre_drift}

As mentioned in the main text, preliminary tests on the deployed system were conducted on a lab-based testing rig in Cambridge.
By utilising delay fibre spools and optical attenuators, the testing rig accurately emulated the primary characteristics of the installed fibres, namely their lengths and losses.
However, as outlined in the Results subsection, some channel properties still differed.
Notably, as discussed in the main text, the polarization drifts introduced by the deployed fibres were significantly lower than those observed in the testing rig.
This difference is attributed to the higher temperature stability of the deployed buried fibres in comparison to the fibre spools utilised in the lab.

\begin{figure}[htb!]
    \centering
    \includegraphics[width=0.8\columnwidth]{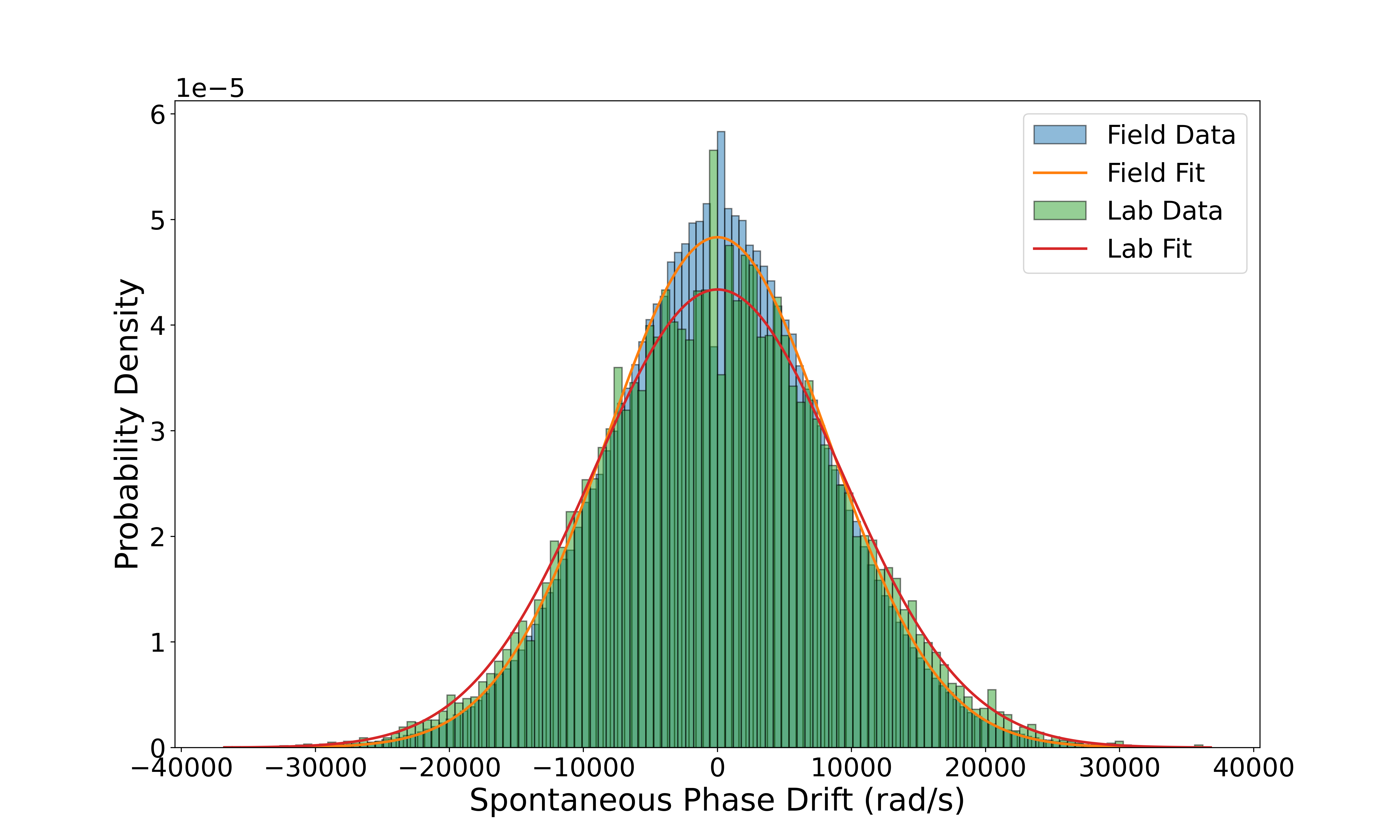}
    \caption{
    Comparison of spontaneous phase drift between field and lab measurements.
    The normalised distribution of the spontaneous phase drift measured in the field (with the lab-based testing rig) is shown in light blue (green).
    Both distributions exhibit a Gaussian shape, consistent with previous observations.
    The normal curves are centred on \qty{0}{\radian\per\second}, with a standard deviation of \qty{\PhaseDriftInLab}{\radian\per\second} for the lab data, and \qty{\PhaseDriftInField}{\radian\per\second} for the field data.
    }
    \label{fig:phase_drift}
\end{figure}

Another relevant property to compare between the two settings is the phase noise.
To explore this, we acquired the free-drifting interference of the quantum signal both in the lab and the field, using our quantum communication prototype and hence the same experimental setup.
From this data, we extracted the distribution of the spontaneous phase drift rate and compared the two cases.
Our analysis, shown in \cref{fig:phase_drift}, reveals that the phase drift rates measured in the field and the lab are very similar.
Specifically, the standard deviation of the spontaneous phase drift measured in the lab is \qty{\PhaseDriftInLab}{\radian\per\second}, while the same quantity in the installed fibres amounts to \qty{\PhaseDriftInField}{\radian\per\second}.
This comparison serves as a valuable reference for the future, providing supporting evidence that the phase noise introduced by fibre spools in the lab effectively simulates that of installed fibres.
It is worth noting that the phase drift measured in our experiment is comparable with the results obtained in previous experiments that used installed optical fibres.

\subsection{The Quantum Encoder}\label{supp:sec:quantum_encoder}
\begin{figure}[tb]
    \centering
    \includegraphics[width=.9\columnwidth]{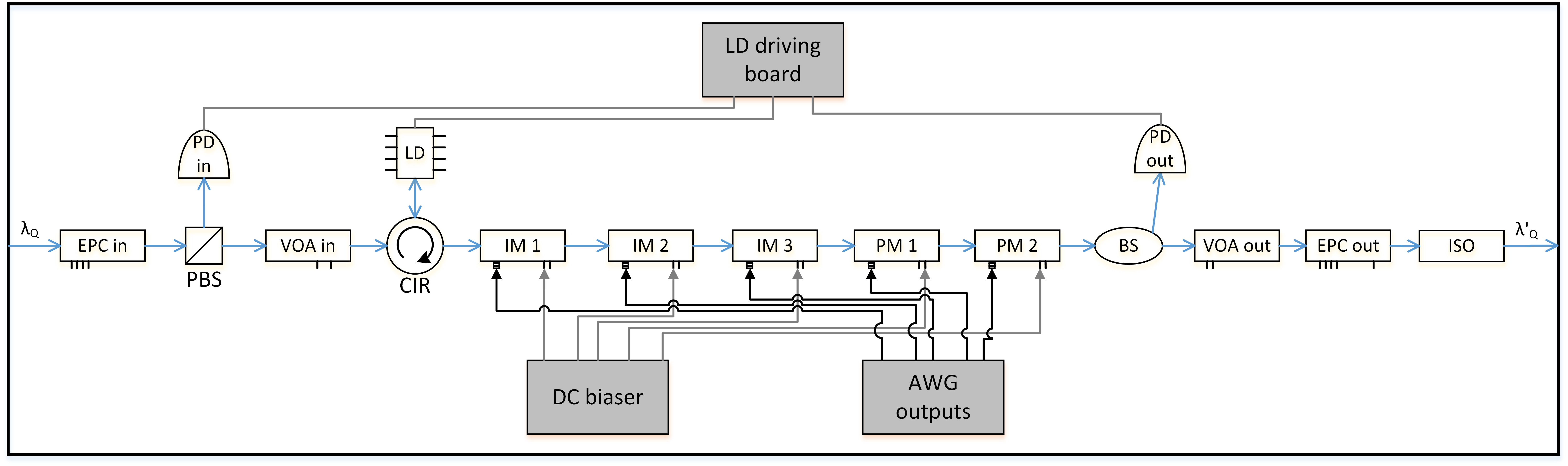}
    \caption{
    Schematic of the Quantum Encoder unit endowed to Alice and Bob.
    The Quantum Encoder receives a \gls*{cw} $\lambda_Q$ signal from the Quantum Signal MUX/DEMUX unit, and it encodes it with the \gls*{tfqkd} protocol.
    The components inside the unit are:
    EPC: Electronic Polarisation Controller;
    PBS: Polarising Beam Splitter;
    PD: Photodiode;
    VOA: Variable Optical Attenuator;
    CIR: Circulator;
    LD: Laser Diode;
    IM: Intensity Modulator;
    PM: Phase Modulator;
    BS: Beam Splitter;
    ISO: Optical Isolator.
    }
    \label{supp:fig:quantum_encoder}
\end{figure}

The Quantum Encoder is the unit in each transmitter node responsible for encoding the quantum communication protocol.
It receives the \gls*{cw} $\lambda_Q$ reference signal distributed by the \textit{service layer} through the Quantum Signal MUX/DEMUX unit.
This signal is processed by several optical modulators/components arranged in series to encode the SNS \gls*{tfqkd} protocol. 
\Cref{supp:fig:quantum_encoder} illustrates the internal components of the Quantum Encoder.

Light entering the Quantum Encoder first encounters an \gls*{epc}, a \gls*{pbs} and a photodiode (PD IN) configured to minimise the optical power recorded by PD IN.
This mechanism compensates for polarisation rotations introduced by $\lambda_Q$ distribution over classical fibres ($F^{AC}_C$ and $F^{BC}_C$), aligning most of the received signals to the \gls*{pbs}'s horizontal polarisation axis.
This mechanism is described in more detail in \cref{supp:sec:stab_mechanisms}, and is one of the several automatic stabilisation routines implemented by the system.

The next set of components includes a \gls*{voa}, followed by an optical \gls*{cir}, with port \#2 connected to a \gls*{ld}.
These components implement an \acrfull*{oil} setup.
The \gls*{voa} attenuates the horizontally polarised component of incoming $\lambda_Q$ exiting the \gls*{pbs} to ensure optimal injection locking of the laser diode cavity.
This way, the laser diode's output is frequency-locked to the $\lambda_Q$ signal, then returned to the \gls*{cir}, and forwarded to ensuing modulators.

Five high-speed electro-optic modulators follow, applying necessary intensity and phase modulations for \gls*{tfqkd} protocol implementation.
Details on the encoded patterns are reported in \cref{supp:sec:encoded_patterns}.
Three intensity modulators (IMs) carve \qty{\PulsesDuration}{\pico\second}-long optical pulses with four different intensities out of the \gls*{cw} signal received by the \gls*{cir}.
\Gls*{im} 1 is driven by a digital \gls*{rf} signal, which carves the received \gls*{cw} signal by driving \gls*{im} 1 between its maximum and minimum transmission settings.
Without further intensity modulation, these carved pulses constitute the brightest pulses produced by the Encoders ($s_A$, $u_A$, and $s_B$).
\Gls*{im} 2 is driven by a three-level \gls*{rf} signal and encodes the decoy states $v_A$, $u_B$, and $v_B$.
\gls*{im} 3 is driven by a digital \gls*{rf} signal, and increases the extinction ratio between the encoded and the vacuum pulses ($w_A$ and $w_B$).

The intensity modulators are followed by two phase modulators.
\Gls*{pm} 1 is driven by an analogue \gls*{rf} signal which encodes an arbitrary phase in the range $\left[ -\frac{\pi}{2}; +\frac{\pi}{2} \right)$.
\Gls*{pm} 2 is driven by a digital \gls*{rf} signal which encodes a phase offset of either $-\frac{\pi}{2}$ or $+\frac{\pi}{2}$.
The combined effect of the two \glspl*{pm} allows the encoding of an arbitrary phase between $\left[ -\pi; +\pi \right)$ onto the transmitted optical pulses.

Each of the five electro-optic modulators receives a dedicated DC biasing signal, optimizing intensity extinction ratio for \glspl*{im} and overall phase offset for \glspl*{pm}.
An \gls*{awg} per transmitter node generates the five \gls*{rf} signals driving the modulators.
These signals are carefully time-synchronised to the \unit{\pico\second}-level for optimal cascaded modulation. 

A \gls*{bs} then redirects most of the light to a second photodiode (PD out), monitoring average optical power.
This photodiode is used to optimise the \gls*{dc} bias of the preceding electro-optic modulators and monitor the average photon flux out of the transmitter.

The last components in the transmitter are \gls*{voa} out, \gls*{epc} out and an optical isolator.
VOA out fine-tunes the intensity of the transmitted optical signal to the target average photon flux of the encoded pattern.
EPC out optimises the polarisation of the transmitted signal to precompensate for polarisation rotations introduced by the quantum channels ($F^{AC}_Q$ and $F^{BC}_Q$).
The mechanism for this precompensation is described in more detail in \cref{supp:sec:stab_mechanisms}.
Finally, to prevent optical back-reflections from entering the Quantum Transmitter through the output port, an optical isolator is installed as the final component of the Quantum Encoder.

At the output of the Quantum Encoder, the signal fully encoded with the quantum communication protocol is referred to as $\lambda'_Q$.

\subsection{Encoding Parameters}\label{supp:sec:encoded_patterns}
In this experiment, we implemented the 4-intensity SNS-TF-QKD quantum communication protocol \cite{Wang.2018, Hu.2019, Xu.2020, Jiang.2020} across a highly asymmetric link.
For the implementation, we used a \EncodedPatternLength-pulse-long repeated encoded pattern.
Interleaved with these are an equal number of phase unmodulated pulses, serving as reference signals to estimate the phase offset between Alice's and Bob's phase reference frames.
The time-averaged outcome of their interference provides the error signal for the \textit{fine} phase stabilisation feedback.

Pattern encoding is executed via two \glspl*{awg}, one per transmitter node, synchronised remotely via an optical \qty{\SysRefClock}{\mega\hertz} reference clock distributed from the central node to transmitter nodes through the classical fibres. 
Detailed information on the \gls*{rf} signals driving the high-speed electro-optic modulators is provided in \cref{supp:sec:quantum_encoder}.

Pattern properties were selected to maximise \gls*{skr} generation under our experimental conditions, considering channel losses and link asymmetry. 
Given the asymmetric configuration for this experiment, the pulses' intensities and sending probabilities are set to satisfy the security condition for asymmetric protocol execution outlined in the \methods.
The chosen pattern properties are summarised in \cref{table:pattern_param}.

Patterns at the transmitters were generated based on probabilities from \cref{table:pattern_param} through fair sampling. 
Initially, a \EncodedPatternLength-elements-long list of pulse pairs matching these probabilities was created. 
Subsequently, the list was randomly shuffled and interleaved with the unmodulated phase referencing pulses.
This approach avoids the effect of statistical fluctuations on pulse pairs distribution inherent in a repeated pattern generated through Monte Carlo approaches.

\begin{table}[!htb]
    \centering
        $\begin{array}{c|cc}
        \toprule
                                        & \text{Alice}      & \text{Bob}    \\
        \midrule
         s\ \textrm{(ph/pulse)}         & 0.52              & 0.24          \\
         u\ \textrm{(ph/pulse)}         & 0.52              & 0.13          \\
         v\ \textrm{(ph/pulse)}         & 0.08              & 0.012         \\
         w\ \textrm{(ph/pulse)}         & 0.0002            & 0.0002        \\
        \midrule
         P_Z\ \textrm{(code\ basis)}    & 80.0\%            & 80.0\%        \\
         P_s (\epsilon)                 & 42.0\%            & 15.0\%        \\
         P_X\ \textrm{(test\ basis)}    & 20.0\%            & 20.0\%        \\
         P_u                            & 5.0\%             & 5.0\%         \\
         P_v                            & 80.0\%            & 80.0\%        \\
         P_w                            & 15.0\%            & 15.0\%        \\
         \bottomrule
        \end{array}$
    \caption{
            Parameters used in implementing the SNS TF-QKD protocol \cite{Wang.2018, Jiang.2020}.
            $s$ denotes the photon flux for signal pulses, while $u$, $v$, and $w$ represent the photon fluxes for decoy states.
            $P_Z$ is the users' probability of choosing the code basis, and $P_s$ ($\epsilon$) is the probability of sending a signal when the $Z$ basis is chosen.
            $P_X$ is the users' probability of encoding a pulse in the test basis, and $P_u$, $P_v$, and $P_w$ are the probabilities of sending the decoy pulses $u$, $v$, $w$.
            }
    \label{table:pattern_param}
\end{table}

\subsection{The Quantum Receiver and Detectors Units}

\begin{figure}[!ht]
    \centering
    \includegraphics[width=0.35\columnwidth]{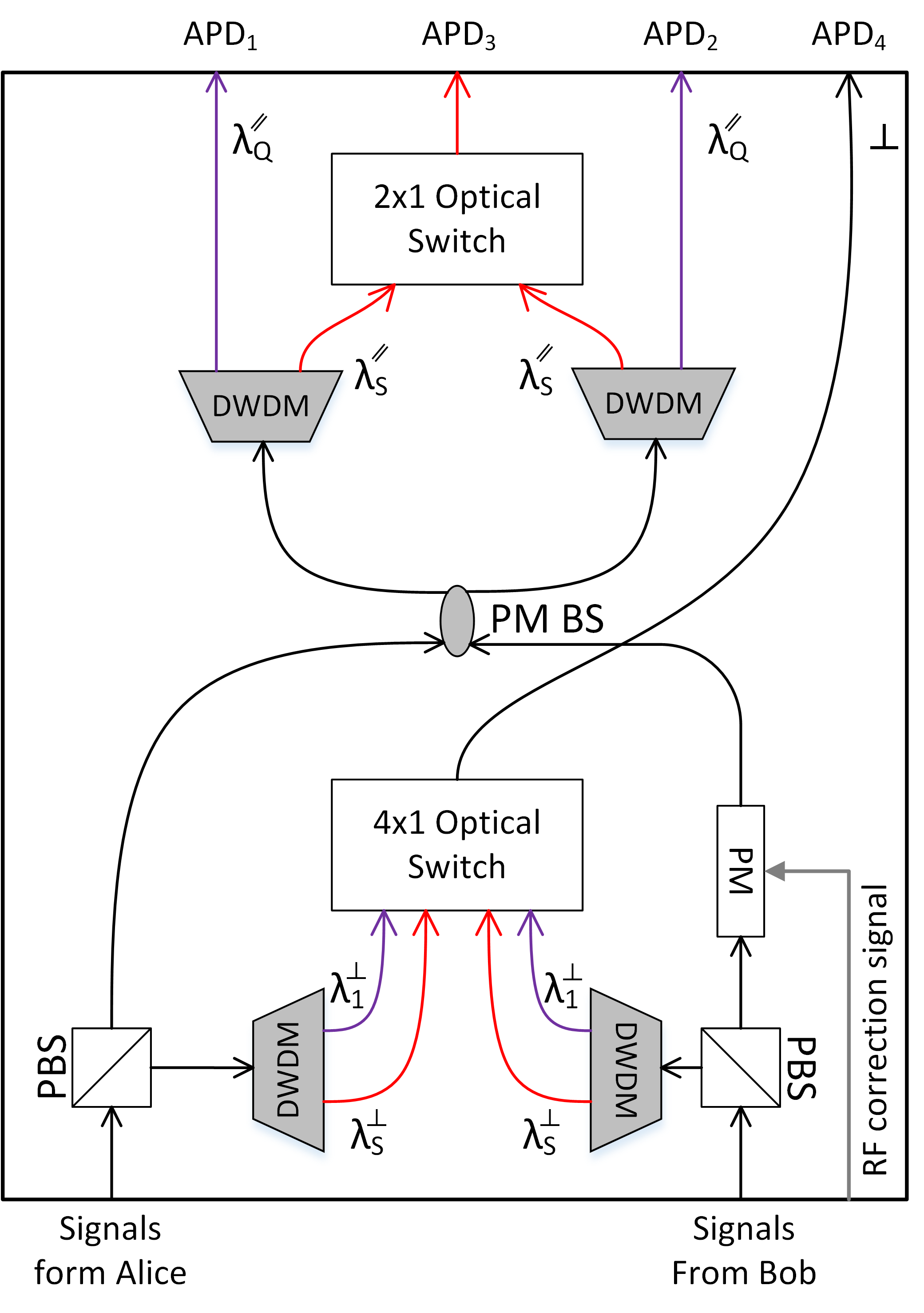}
    \caption{
    Schematic of the Quantum Receiver unit.
    PM BS: Polarisation Maintaining Beam Splitter,
    PM: Phase Modulator,
    PBS: Polarising Beam Splitter,
    DWDM: Dense Wavelength Demultiplexer,
    APD: Avalanche Photodiode.
    }
    \label{supp:fig:quantum_receiver}
\end{figure}

The Quantum Receiver combines signals received from Alice and Bob and extracts their orthogonal polarisation components, which are induced by polarisation rotations over the quantum channel.
It then directs all the signals to four \glspl*{apd} for detection. 
Inputs are from the $F^{AC}_Q$ and $F^{BC}_Q$ fibres linking Charlie to Alice and Bob, respectively.
Inside the receiver, signals first pass through two \glspl*{pbs} to align the polarisation of the incoming signals to the polarisation axes of the ensuing optical components.
Vertical polarization components are passed to a 4x1 optical switch, which dynamically selects and directs one to APD\textsubscript{4} for polarisation optimisation.
The horizontal polarisation components of Bob's signals pass through a \gls*{pm}, which serves as an actuator for the \textit{coarse} feedback mechanism.
The horizontal polarisation components of Alice's and Bob's signals then interfere on a polarisation-maintaining \gls*{bs}.
The interference outcomes are frequency demultiplexed by two \glspl*{dwdm}, with the outputs at $\lambda_Q$ directed to APD\textsubscript{1} and APD\textsubscript{2}.
Outputs at $\lambda_S$ are dynamically selected by a 2x1 optical switch and are directed to APD\textsubscript{3}.
\Cref{supp:fig:quantum_receiver} for a schematic of the quantum receiver, and \cref{supp:tab:quant_receiver} for component characterisation.

\begin{table}[!ht]
    \centering
        \begin{tabular}{lc}
            \toprule
                Component           & Transmission  \\
            \midrule
                PBSs                & 93.0 \%       \\
                PM BSs              & 93.7 \%       \\
                DWDMs               & 87.3 \%       \\
                Phase Modulator     & 64.6 \%       \\
            \bottomrule
        \end{tabular}
    \caption{
    Transmissivity of the optical components inside the quantum receiver.
    The overall transmission of the receiver (excluding the \gls*{pm}) is measured to \qty{76.0}{\percent}.
    }\label{supp:tab:quant_receiver}
\end{table}

The APDs 1 to 4 are \qty{-30}{\celsius}-cooled \qty{1}{\giga\hertz}-gated self-differenced \glspl*{apd}~\cite{Yuan.2008}.
Critical to the successful execution of the quantum communication protocol is the selection of deadtime on time tagger used to analyse the \glspl*{apd} output signals.
For \glspl*{apd} 1 and 2, which are used to monitor interference outcomes of protocol-encoded pulses, a long deadtime of \qty{\DeadTimeWindow}{\micro\second} was set to effectively suppress afterpulse effect on secret key rate generation~\cite{Hadfield.2009, Zhang.2015, Humer.2015}.
Conversely, for \glspl*{apd} 3 and 4, responsible for monitoring the signals supporting the protocol execution, a much shorter deadtime of \qty{64}{\nano\second} was chosen to maximise the maximum countrate on these detectors. 
The characterisation of all \glspl*{apd} is reported in \cref{supp:tab:apds}.

\begin{table}[!ht]
    \centering
        \begin{tabular}{lc@{\hskip 0.2in}c}
            \toprule
                                    & APD\textsubscript{1} \& APD\textsubscript{2}      & APD\textsubscript{3} \& APD\textsubscript{4}      \\
            \midrule
                Efficiency          & 14.5 \%                                           & 15 \%                                             \\
                Dark counts         & \qty{450}{\hertz}                                 & \qty{500}{\hertz}                                 \\
                Set dead-time       & \qty{\DeadTimeWindow}{\micro\second}              & \qty{64}{\nano\second}                            \\
            \bottomrule
        \end{tabular}
    \caption{
    Summary of the properties of the APDs used in the experiment.
    }\label{supp:tab:apds}
\end{table}

\subsection{Stabilisation Mechanisms}\label{supp:sec:stab_mechanisms}
\begin{figure}[ht!]
    \centering
    \includegraphics[width=1.15\columnwidth,height=\textheight,keepaspectratio,angle=90]{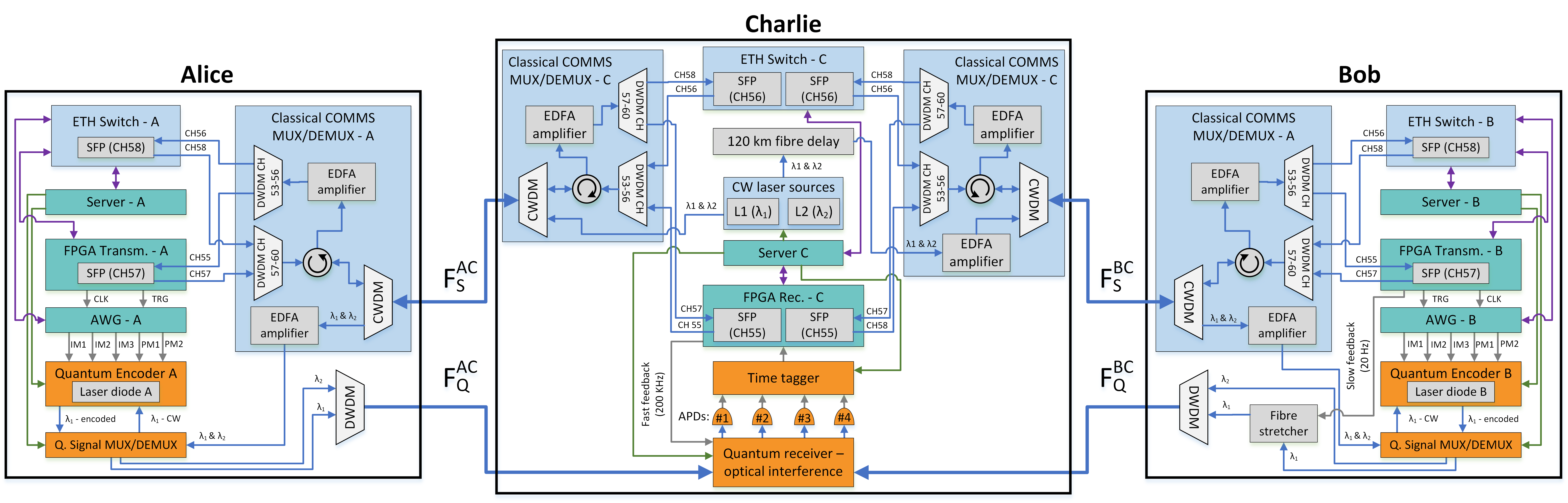}
    \caption{
    Detailed representation of the developed coherent quantum communication system.
    }
    \label{fig:supp:full_sys}
\end{figure}

To support the operation of the system, which ran continuously for \qty{\ProtExecTime}{\hour} to generate the key material that led to the positive \gls*{skr} generation, several automatic optimisation mechanisms were implemented and are described below.
\begin{description}
    \item[Time synchronisation of high-speed electronics]
    Time synchronization of high-speed electronics (\glspl*{awg} at the transmitters and time tagger at the receiver node) is achieved through a custom-developed system. 
    Pairwise interconnected \glspl*{fpga} at the nodes support the precise time synchronization.
    Charlie's FPGA generates a \qty{\SysRefClock}{\mega\hertz} local reference clock, which is distributed optically to Alice and Bob via SFPs and the classical fibres $F^{AC}_C$ and $F^{BC}_C$.
    Fast PLLs synchronise the FPGAs, ensuring coordinated operation of all equipment.
    RF reference clocks and trigger signals from each FPGA synchronise the AWGs and time tagger.
    
    \item[Optimisation of the polarisation for the Quantum Encoder]
    The $\lambda_Q$ reference signals, distributed from Charlie to Alice and Bob, undergo polarisation rotations introduced by the classical fibres $F^{AC}_C$ and $F^{BC}_C$.
    To counter these rotations, each Quantum Encoder features a polarisation optimisation subsystem.
    This subsystem, comprised of the EPC in, PBS, and PD in components at the entrance of the Quantum Encoder(represented in \cref{supp:fig:quantum_encoder}), continuously adjusts the settings of EPC in using an automatic algorithm. 
    The algorithm minimises the power readout at PD in, ensuring alignment of the incoming light with the horizontal polarization axis of the PBS and compensating for the polarisation rotations introduced by the classical fibres.
    
    \item[OIL optimisation] 
    The \gls*{oil} parameters for both transmitters are initially optimised in the lab to ensure optimal coherence transfer between the primary (L\textsubscript{S}) and secondary (LD) laser sources.
    This optimization includes adjusting parameters such as laser diode temperature, DC bias, and injected optical power. 
    Upon commencing the experiment, these parameters are replicated in the field.
    The injected power is fine-tuned to the optimal level by adjusting the \gls*{voa} in (see \cref{supp:fig:quantum_encoder}), guided by the power reading from a photodiode within the \gls*{ld} cavity.
    Once the injected power reaches the optimal level, the DC bias and temperature of the \gls*{ld} are adjusted to the desired settings using a \gls*{ld} driving board (as depicted in \cref{supp:fig:quantum_encoder}).
    Thanks to the stability of the reference $\lambda_Q$ signal transmitted by the service layer and the efficacy of the polarisation optimisation routine for the incoming $\lambda_Q$ signal, the \gls*{oil} optimisation process only needs to be performed once at the beginning of the experiment.
 
    \item[DC bias optimisation for the high-speed electro-optic modulators]
    Initial DC bias optimisation of the electro-optic modulators, especially \glspl*{im}, is executed at the beginning of the experiment by scanning the DC bias applied to each modulator while concurrently monitoring the transmitted power using PD out  (see \cref{supp:fig:quantum_encoder}).
    This procedure is conducted individually for each \gls*{im} while the other \glspl*{im} are set to pass-all configuration and enables the determination of the starting optimal DC offset for each \glspl*{im}.
    During the protocol execution, an additional mechanism is employed to mitigate the spontaneous \gls*{dc} drift of the modulators.
    This mechanism relies on monitoring the extinction ratio between the various types of encoded pulses transmitted by one user when the other user is sending vacuum states.
    From these measurements made by Charlie, each user can tune the DC biases of their \glspl*{im} to maintain the desired intensity relation between the pulses.
    
    \item[Optimisation of the photon flux out of the transmitter nodes] 
    Optimisation of the average intensity of the encoded patterns just out of the transmitter nodes to align them with the photon fluxes specified in \cref{table:pattern_param} is done through the power readout of PD out, and the attenuation of the outgoing signal through VOA out.
    During the testing phase of the system, the intensity ratio between the signal output recorded by PD out (see \cref{supp:fig:quantum_encoder}) and the signal injected into the quantum fibre is carefully characterised. 
    This ratio remains fixed and can only be adjusted by additional attenuation provided by VOA out.
    As a result, the power measured at PD out serves as an indicator of the average optical intensity of the $\lambda'_Q$ signal produced by the transmitter node. 
    The PD out readout, alongside the knowledge of the encoded pattern properties, enables Alice and Bob to set their photon fluxes in alignment with those specified in \cref{table:pattern_param}, ensuring adherence to the security requirements of the protocol implementation. 
    The power readout of PD out is continuously monitored throughout the protocol execution, and the attenuation of VOA out is fine-tuned to maintain consistency with the photon fluxes outlined in \cref{table:pattern_param}.
    
    \item[Optimisation of $\lambda'_Q$ and $\lambda'_S$ polarisation]
    Similarly to the polarisation rotations introduced by the classical fibres, which are addressed by the polarisation stabilisation routine implemented by the Quantum Encoder, the polarisation rotation introduced by the quantum fibres $F^{AC}_Q$ and $F^{BC}_Q$ also needs to be compensated for. 
    In this case, compensation is applied at the transmitting end of the fibre rather than the receiving end.
    This configuration is chosen to minimise losses at the quantum receiver and enhance protocol performance. 
    Precompensation is applied by the EPC out in \cref{supp:fig:quantum_encoder} for $\lambda'_Q$ and by an EPC in the Quantum Signal MUX/DEMUX unit (see\cref{fig:supp:full_sys}) for $\lambda'_S$.

    The polarisation optimisation mechanism adjusts the polarisation of all signals reaching the Quantum Receiver by minimising in turn the number of counts recorded by APD 4 for each signal directed to it by the 4x1 optical switch (see \cref{supp:fig:quantum_encoder}).
    
    \item[Off-band phase stabilisation]
    Due to the critical role played by the off-band phase stabilisation in the protocol execution, a detailed description of this mechanism is outlined in the Methods subsection of the manuscript. 
    
\end{description}

\clearpage
\subsection{Detailed Schematic of the Quantum Communication System}

\clearpage
\subsection{Detailed Experimental Results for SNS TF-QKD Protocol}
\Cref{table:exp_results_TWCC_fs} summarises the detailed experimental results obtained from the implementation of the SNS \gls*{tfqkd} coherent quantum communication protocol between Frankfurt and Kehl. 
\textit{N total sent} is the total number of encoded quantum pulses sent throughout the protocol execution. 
The number of pulses sent in each Alice-Bob pulses pair configuration can be calculated by multiplying the respective configuration probabilities (deducible from \cref{table:pattern_param}) by \textit{N total sent}.
The number of pulses detected in these configurations are reported in \cref{table:exp_results_TWCC_fs} with format ``Detected-B1B2t1t2'', where B1 and t1 (B2 and t2) are the basis and the type of pulse sent by Alice (Bob).

\begin{table}[!ht]
    \centering
    \resizebox{0.5\textwidth}{!}{%
        $
        \begin{array}{c|c}
            \toprule
                \text{Fibre Length (km)} & 253.9 \\
            \midrule
                \text{Alice - Charlie (km)} & 156.7 \\
                \text{Bob - Charlie (km)} & 97.2 \\
            \midrule
                \text{N total sent} & 1.36581\times 10^{13} \\
                \text{Phase mismatch acceptance} & 22.5{}^{\circ} \\
            \midrule
                \text{Z error rate (before AOPP)} & \text{30.7$\%$} \\
                \text{Z error rate (after AOPP)} & \text{3.56$\%$} \\
                \text{Xuu error rate} & \text{4.61$\%$} \\
                \text{Xvv error rate} & \text{5.83$\%$} \\
                \text{Phase error rate} & \text{5.08$\%$} \\
            \midrule
                \text{Detected-ZZss} & 80342420 \\
                \text{Detected-ZZsn} & 104894262 \\
                \text{Detected-ZZns} & 86381667 \\
                \text{Detected-ZZnn} & 4163565 \\
                \text{Detected-ZXsu} & 4346359 \\
                \text{Detected-ZXsv} & 28080211 \\
                \text{Detected-ZXsw} & 4647897 \\
                \text{Detected-ZXnu} & 3924498 \\
                \text{Detected-ZXnv} & 3299312 \\
                \text{Detected-ZXnw} & 180031 \\
                \text{Detected-XZus} & 2405448 \\
                \text{Detected-XZun} & 3107519 \\
                \text{Detected-XZvs} & 31252233 \\
                \text{Detected-XZvn} & 8882151 \\
                \text{Detected-XZws} & 5576883 \\
                \text{Detected-XZwn} & 261425 \\
                \text{Detected-XXuu} & 129224 \\
                \text{Detected-XXuv} & 841061 \\
                \text{Detected-XXuw} & 136832 \\
                \text{Detected-XXvu} & 1467103 \\
                \text{Detected-XXvv} & 3873980 \\
                \text{Detected-XXvw} & 392002 \\
                \text{Detected-XXwu} & 249794 \\
                \text{Detected-XXwv} & 418392 \\
                \text{Detected-XXww} & 11671 \\
             \midrule
                 \text{SKC0 (bit/channel use)} & 3.62 \cdot 10^{-6} \\
                 \text{SKC0 (bit/s)} & 1812.0 \\
            \midrule
                \text{Number of secure bits generated (bits)} & 3.01 \cdot 10^6 \\
                \text{SKR (bit/signal)} & 2.20 \cdot 10^{-7} \\
                \text{SKR (bit/s)} & 110.1 \\
             \bottomrule
        \end{array}
        $
        }
    \caption{Detailed results obtained from the SNS TF-QKD protocol implementation between Frankfurt and Kehl.}
    \label{table:exp_results_TWCC_fs}
\end{table}

\end{document}